%% file: msppb-preprint.tex
\documentclass[ALICE,manyauthors]{cernphprep}
\input{commands.tex}  
\usepackage{svn-multi}  
\svnidlong   
{$HeadURL: svn+ssh://dialexan@svn.cern.ch/reps/msppb/paper/msppb-preprint.tex $} 
{$LastChangedDate: 2015-11-15 20:50:53 +0000 (Sun, 15 Nov 2015) $} 
{$LastChangedRevision: 226 $} 
{$LastChangedBy: dialexan $}
\svnid
{$Id$}

\usepackage[usenames]{color}
\usepackage{multirow}

\RequirePackage{color}\definecolor{RED}{rgb}{1,0,0}\definecolor{BLUE}{rgb}{0,0,1}

\def\includeBW{1}

\usepackage[comma,square,numbers,sort&compress]{natbib}
\usepackage{hyperref}
\usepackage{lineno}

\begin{document}%

\begin{titlepage}
\PHyear{2015}
\PHnumber{327}      
\PHdate{18 December}  
%

\title{Multi-strange baryon production in p--Pb collisions at $\mathbf{\sqrt{{\textit s}_{\mathbf NN}}}=5.02$ TeV}
\ShortTitle{Multi-strange baryon production in p--Pb collisions at \snn~=~5.02~TeV}   

\Collaboration{ALICE Collaboration\thanks{See Appendix~\ref{app:collab} for the list of collaboration members}}
\ShortAuthor{ALICE Collaboration} 

\begin{abstract}
\input{abstract.tex}
\end{abstract}
\end{titlepage}
\setcounter{page}{2}

\input{msppb-main.tex}               
%
%

\newenvironment{acknowledgement}{\relax}{\relax}
\begin{acknowledgement}
\section*{Acknowledgements}
\input{acknowledgements.tex}    
\end{acknowledgement}

\bibliographystyle{utphys}   
\bibliography{msppb}

\newpage
\appendix
\section{The ALICE Collaboration}
\label{app:collab}
\input{Alice_Authorlist_2015-Nov-27.tex}  
\end{document}

%% file: commands.tex
\newcommand{\Om}{$\Omega^-$}
\newcommand{\Mo}{$\overline{\Omega}^+$}
\newcommand{\X}{$\Xi^-$}
\newcommand{\Ix}{$\overline{\Xi}^+$}
\newcommand{\Xis}{$\Xi^{\pm}$}
\newcommand{\Oms}{$\Omega^{\pm}$}

\newcommand{\GeVc}{GeV/$c$}

\newcommand{\GeVmass}{GeV/$c^2$}
\newcommand{\MeVmass}{MeV/$c^2$}
\newcommand{\allpart}{\X, \Ix, \Om and \Mo}
\newcommand{\thermus} {\rm{THERMUS}}

\newcommand{\VZERO}        {\rm{V0}~}
\newcommand{\VZEROA}       {\rm{V0A}~}
\newcommand{\VZEROC}       {\rm{V0C}~}
\newcommand{\Vdecay} 	 {$V^{0}$~}
\newcommand{\pip}          {$\pi^{+}$}
\newcommand{\pim}          {$\pi^{-}$}

\newcommand{\pbar}         {$\rm\overline{p}$}

\newcommand{\lmb}          {\ensuremath{\Lambda}}
\newcommand{\almb}         {\ensuremath{\bar{\Lambda}}}

\newcommand{\pp}           {pp}

\newcommand{\PbPb}         {\mbox{Pb--Pb}}
\newcommand{\pPb}          {\mbox{p--Pb}}
\newcommand{\AuAu}         {\mbox{Au--Au}}

\newcommand{\dNdeta}       {\ensuremath{\mathrm{d}N_\mathrm{ch}/\mathrm{d}\eta}}

\newcommand{\s}            {\ensuremath{\sqrt{s}}}
\newcommand{\pt}           {\ensuremath{p_{\rm T}}}

\newcommand{\hlab}         {\ensuremath{\eta_{\rm lab}}}

\newcommand{\ycms}         {\ensuremath{y_{\rm CMS}}}

\newcommand{\snn}          {\ensuremath{\sqrt{s_{\rm NN}}}}

\newcommand{\avbT}         {\ensuremath{\left< \mathrm{\beta}_{\rm T}\right>}}

\newcommand{\avg}[1]       {\ensuremath{\left\langle#1\right\rangle}}



%% file: abstract.tex
The multi-strange baryon yields in \PbPb~collisions have been shown to exhibit an enhancement relative to \pp~reactions. In this work, $\Xi$ and $\Omega$ production rates have been measured with the ALICE detector as a function of transverse momentum, \pt,  in p\nobreakdash--Pb collisions at a centre-of-mass energy of \snn~= 5.02 TeV. The results cover the kinematic ranges 0.6 \GeVc\ $<$ \pt\ $<$ 7.2 \GeVc\ and 0.8 \GeVc\ $<$ \pt\ $<$ 5 \GeVc, for  $\Xi$ and $\Omega$ respectively, in the common rapidity interval -0.5 $<\ycms<$ 0. Multi-strange baryons have been identified by reconstructing their weak decays into charged particles. The \pt\ spectra are analysed as a function of event charged--particle multiplicity, which in \pPb\ collisions ranges over one order of magnitude and lies between those observed in pp and \PbPb\ collisions.\if\includeBW1 The measured \pt\ distributions are compared to the expectations from a Blast-Wave model. The parameters which describe the production of lighter hadron species also describe the hyperon spectra in high multiplicity \pPb\ collisions\fi.  The yield of hyperons relative to charged pions is studied and compared with results from \pp~and \PbPb~collisions. A statistical model is employed, which describes the change in the ratios with volume using a canonical suppression mechanism, in which the small volume causes a species-dependent relative reduction of hadron production.  The calculations, in which the magnitude of the effect depends on the strangeness content, show good qualitative agreement with the data. 


%% file: msppb-main.tex
 \svnidlong 
{$HeadURL: svn+ssh://svn.cern.ch/reps/alipap/id1537/paper/msppb-main.tex $} 
{$LastChangedDate: 2015-12-22 14:34:00 +0100 (Tue, 22 Dec 2015) $} 
{$LastChangedRevision: 1678 $} 
{$LastChangedBy: dialexan $}
 \svnid
 {$Id: msppb-main.tex 245 2015-12-18 20:59:08Z dialexan $}

\section{Introduction}

Collisions of heavy nuclei at ultra-relativistic energies allow the study of a deconfined state of matter, the Quark-Gluon Plasma, in which the degrees of freedom
are partonic, rather than hadronic. The role of strange hadron yields
in searching for this state was pointed out at an early stage
\cite{PhysRevLett.48.1066}. It was subsequently
found that in high energy nucleus-nucleus (A--A) collisions at the
Super Proton Synchrotron (SPS), the Relativistic Heavy Ion Collider (RHIC)
and the Large Hadron Collider (LHC) the abundances of strange and
multi-strange baryons are compatible with those from thermal
statistical model calculations
\cite{Andersen1998209,Andersen1999401,Afanasiev2002275,Antinori200468,PhysRevLett.93.022302,PhysRevLett.92.182301,PhysRevLett.98.062301,PhysRevC.77.044908,PbPbmsALICE}. 

In smaller collision systems at the same centre-of-mass energies, in
particular proton-proton (pp) collisions, the relative abundance of
multi-strange baryons is lower with respect to A--A collisions,
whether normalised to participant nucleons or produced particles
(pions or charged hadrons). This led to the interpretation that
strangeness enhancement is observed in A--A collisions. Attempts to
explain this phenomenon include the application of a canonical formalism in the statistical model,
replacing the grand canonical approach, in which the requirement to
conserve the strangeness quantum number when producing (multi-)strange
baryons in small systems is imposed \cite{Redlich:2002dm}. This means
that strange hadrons are produced with a lower relative abundance in small systems, an effect known as canonical suppression.  Such a theoretical
framework has been used to make predictions for LHC energies
\cite{Kraus:2009iv}. Further complications in the interpretation arise
when the produced system, although small, is formed in peripheral A--A
collisions where the particle production may not be from a contiguous
volume due to core-corona effects 
\cite{Becattini:2008vj,Aichelin:2008mi}.
Evidence
for this effect was seen at RHIC where a canonical suppression
calculation based on the estimated number of participant nucleons
could not successfully reproduce the data \cite{Agakishiev:2012jv}. A
cleaner way to investigate canonical suppression effects is provided by proton--nucleus (p--A) 
collisions.

Proton--nucleus collisions provide an opportunity to study the
\pt-dependence of the particle spectra created in a system with a
different, more compact, initial geometry than A--A collisions where a
similar number of charged particles are produced. Studying this
dependence is important in determining the applicability of
hydrodynamics \cite{Schnedermann:1993ws} which has been successful in describing the
particle spectra in A--A collisions \cite{PhysRevC.90.054912,Melo:2015iva,Adams2005102}.

At the LHC the combination of the rise in particle production per
nucleon-nucleon collision with increasing \s~ and a dedicated
\pPb~data-taking period have enabled the ALICE experiment to collect a
large sample of \Xis\ and \Oms. In this Letter, we set out the methods
for these studies, present the results obtained and discuss how they fit into
a theoretical picture. 

\section{Sample and data analysis}

The results presented in this Letter were obtained from a sample of the data
collected with the ALICE detector \cite{1748-0221-3-08-S08002} during the LHC \pPb\ run at \snn~=~5.02~TeV 
in the beginning of 2013. The two scintillator arrays \VZEROA
(direction of Pb beam), and \VZEROC (direction of p beam), covering pseudo-rapidity ranges of
$2.8<\eta<5.1$ and -3.7$<\eta<$-1.7,  respectively, served both as triggering detectors and for determining the event multiplicity
class \cite{1748-0221-8-10-P10016}. The tracking of particles in the central barrel, covering $|\eta|<0.9$, takes place in the Inner Tracking System
(ITS), which consists of the two innermost silicon pixel layers,
surrounded by two silicon drift and two silicon strip layers, all
placed within a radius of 43 cm, and the Time Projection Chamber (TPC),
a large cylindrical drift chamber filled with a Ne-CO$_{2}$ gas mixture
\cite{1748-0221-3-08-S08002}. Measurements of the energy loss by
charged particles in the gas allow particles to be identified with this detector. 

A trigger requiring a coincidence within less than 1 ns in the \VZERO
detectors selected around 100 million events, which are mainly
non-single diffractive (NSD) events and contain a negligible
contribution from single diffractive (SD) and electromagnetic (EM)
processes \cite{ALICE_PseudoRapDensitypPb}. A dedicated
radiator-quartz detector (T0) provided a measurement of the event time
of the collisions. The V0 and T0 time resolutions allowed discrimination of
beam-beam interactions from background events in the interaction
region. Further background suppression was applied in the offline
analysis using time information from two neutron Zero Degree
Calorimeters (ZDC), as was performed in previous \pPb\ analyses
\cite{piKpLambda_pPb}. Primary vertices (PVs) were selected if their
position along the beam axis was reconstructed within 10 cm of the
geometrical centre of the detector. In Monte Carlo (MC) studies 
an efficiency of 99.2$\%$ for this trigger was obtained, while the joint
trigger and primary vertex reconstruction efficiency lies at 97.8$\%$
\cite{ALICE_PseudoRapDensitypPb}. The estimated mean number of
interactions per bunch crossing was below 1$\%$ in the sample chosen
for this analysis.  

The analysed events were divided into seven multiplicity percentile classes according
to the total number of particles measured in the forward \VZEROA detector. The
efficiency-corrected mean number of charged primary particles per unit
rapidity ($\avg{\dNdeta}$) within $-0.5<\eta<0.5$ in the laboratory reference
frame for each of these multiplicity bins were published in~\cite{piKpLambda_pPb}. 

Due to the asymmetric energies of the proton and lead ion beams, a
consequence of the 2-in-1 magnet design of the LHC, the
nucleon-nucleon centre-of-mass system is shifted by
0.465 units of rapidity in the direction of the proton beam with
respect to the laboratory frame.  The measurements reported in this
Letter were performed in the central rapidity window defined in the
centre-of-mass frame within $-0.5<y<0$, where negative rapidity corresponds to the side of
the detector into which the Pb beam travels.

The identification of multi-strange baryons was based on the
topology of their weak decays through the reconstruction of the tracks
left behind by the decay products, referred to as the daughter particles. The
daughters of the
$\Xi^{-}\rightarrow\Lambda\pi^{-}$ (BR: 99.9$\%$),
$\Omega^{-}\rightarrow\Lambda K^{-}$  (BR: 67.8$\%$)  and
the subsequent $\Lambda\rightarrow p\pi^{-}$(BR: 63.9$\%$) weak decays
\cite{Agashe:2014kda}, as
well as the corresponding decays of the $\overline{\Xi}^{+}$ and $\overline{\Omega}^{+}$, were
reconstructed by combining track information from the TPC and the
ITS \cite{ms7ppALICE}. Proton, anti-proton and charged $\pi$ and K tracks were identified in the TPC via their measured
energy deposition, which was compared with a mass-dependent
parameterisation of ionisation loss in the TPC gas as a
function of momentum  \cite{TheALICECollaboration:2014kr}. All daughter candidates were
required to lie within 4$\sigma$ of
their characteristic Bethe-Bloch energy loss curve. Multi-strange
candidates were selected through the geometrical association of the \Vdecay
component (\lmb~or \almb~decay) to a further secondary, `bachelor'
track (identified as $\pi^\pm$ or K$^\pm$). In this process, several
geometrical variables were measured for each candidate,
and criteria were set on them in order to purify the
selected sample: numerical values for the selection cuts applied are reported in Table \ref{tCuts}. These selections are similar
to those in the pp measurements \cite{ms7ppALICE}, a consequence of
the low multiplicities present in the detector in the \pPb\
collisions. As a result the correction factors for the efficiency are
also similar. In addition to the settings on topological variables, a
cut has been applied on the \Vdecay invariant mass window of
$\pm$8 \MeVmass~from the nominal
$\Lambda$ mass \cite{Agashe:2014kda}. Further restrictions were set on the proper lifetime
of the \Xis\ and \Oms. By requiring this variable to be less than
3 times the mean decay length (4.91 cm and 2.46 cm, respectively), we
discarded low-momentum secondary particles and false multi-strange
candidates, the daughter tracks of which originated from interactions
with detector material. 

\begin{table}[!tb]
\begin{center}
\begin{tabular}{ll}
\hline
\noalign{\smallskip}
 \multicolumn{2}{l}{\bf ~~~\Vdecay finding criteria}                      \\
\noalign{\smallskip}
\hline
\noalign{\smallskip}
 \it DCA: h$^\pm$ to PV                        & $>0.04$ (0.03) cm                           \\ 
 \it DCA: h$^-$ to h$^+$                       & $<1.5$ standard deviations                  \\
 \it $\Lambda$ mass ($m_{\mathrm{V0}}$)         & $1.108<m_{\mathrm{V0}}<$ 1.124 \GeVmass     \\
 \it Fiducial volume (R$_{\rm{2D}}$)                 & $R_{\rm{2D}} >$ 1.1 (1.2) cm                 \\
 \it \Vdecay pointing angle                          & $\cos\theta_{\mathrm{V0}}>$ 0.97                     \\

\noalign{\smallskip}
\hline
\noalign{\smallskip}
\multicolumn{2}{l}{\bf ~~~Cascade finding criteria}                  \\
\hline
\noalign{\smallskip} 
  \it Proper decay length       & $<3 \times$ mean decay length\\
  \it DCA: $\pi^{\pm}$ (K$^{\pm}$) to PV       & $>0.04$ cm\\
  \it DCA: \Vdecay to PV                            & $>0.06$ cm\\
  \it DCA: $\pi^{\pm}$ (K$^{\pm}$) to \Vdecay       & $< 1.3$ cm\\    
  \it Fiducial volume (R$_{\rm{2D}}$)                & $R_{\rm{2D}}>$ 0.5 (0.6) cm\\
  \it Cascade pointing angle                    & $\cos\theta_{\rm{casc}}>$ 0.97\\
\noalign{\smallskip}                     
\hline 
\end{tabular}
\end{center}
\caption{The parameters for \Vdecay (\lmb~and \almb) and cascades
  (\Xis\ and \Oms) selection criteria. Where a criterion for \Xis\ and \Oms\ finding differs, the value for the \Oms\ case is in parentheses.
DCA represents ``distance of closest approach," PV the primary vertex,
$\theta$ is the angle between the momentum vector of the reconstructed \Vdecay or cascade, and the displacement vector between the decay and primary vertices. The curvature of the cascade particle's trajectory is neglected.}
\label{tCuts}
\end{table}

\begin{figure*}[t!]
  \begin{flushleft}
    \includegraphics[width=0.495\textwidth]{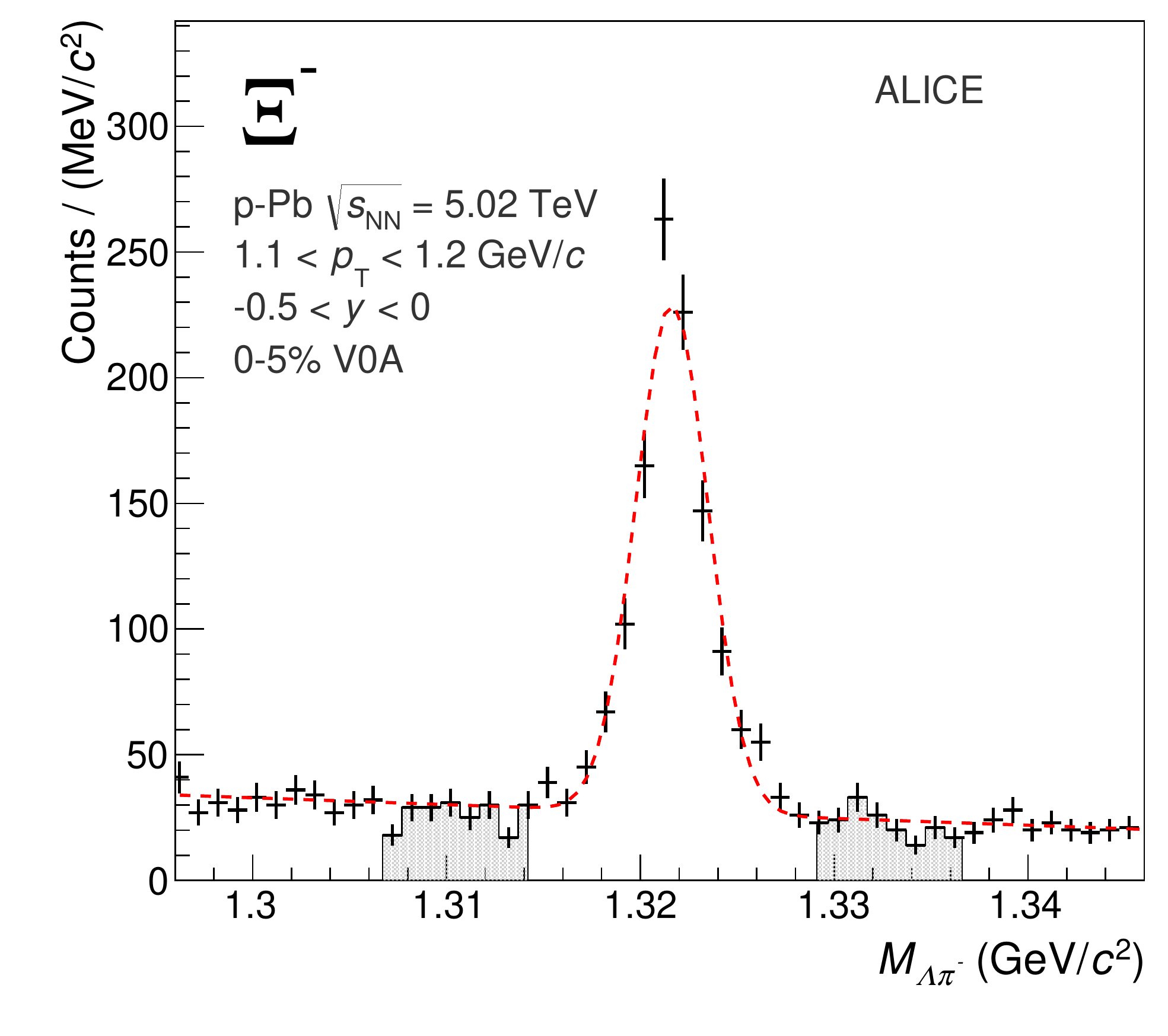}
    \includegraphics[width=0.495\textwidth]{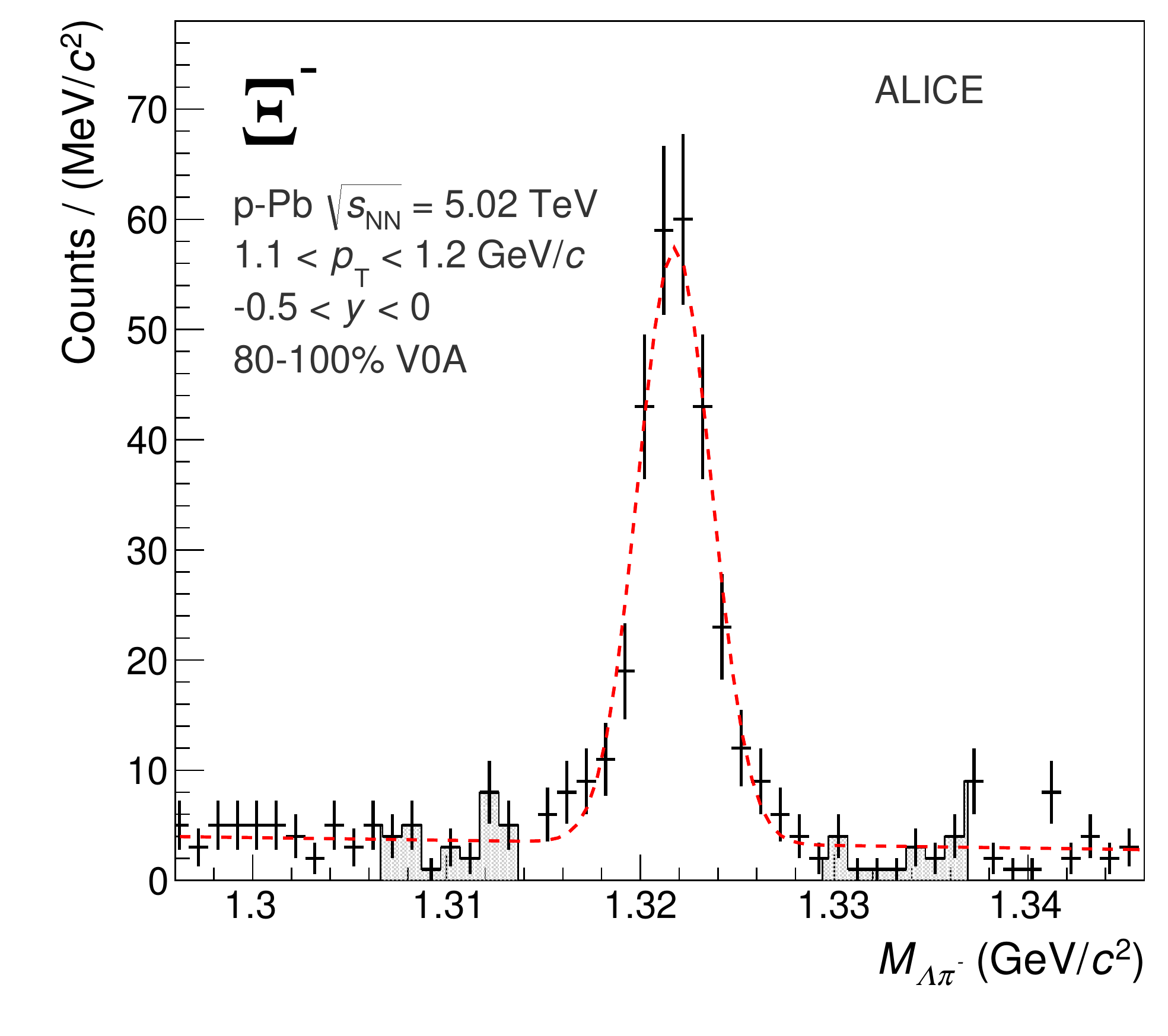} \\
    \includegraphics[width=0.495\textwidth]{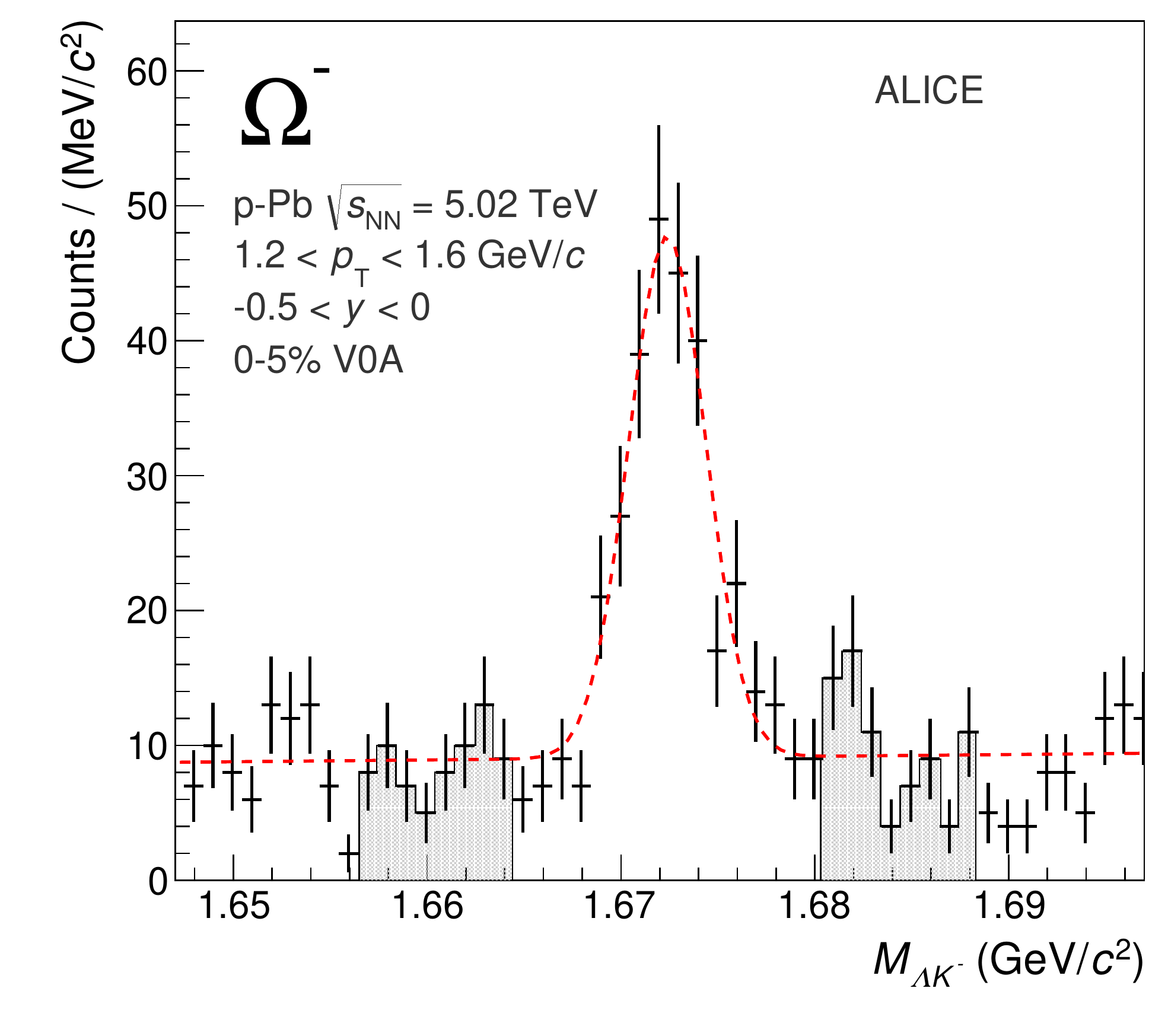}
    \includegraphics[width=0.495\textwidth]{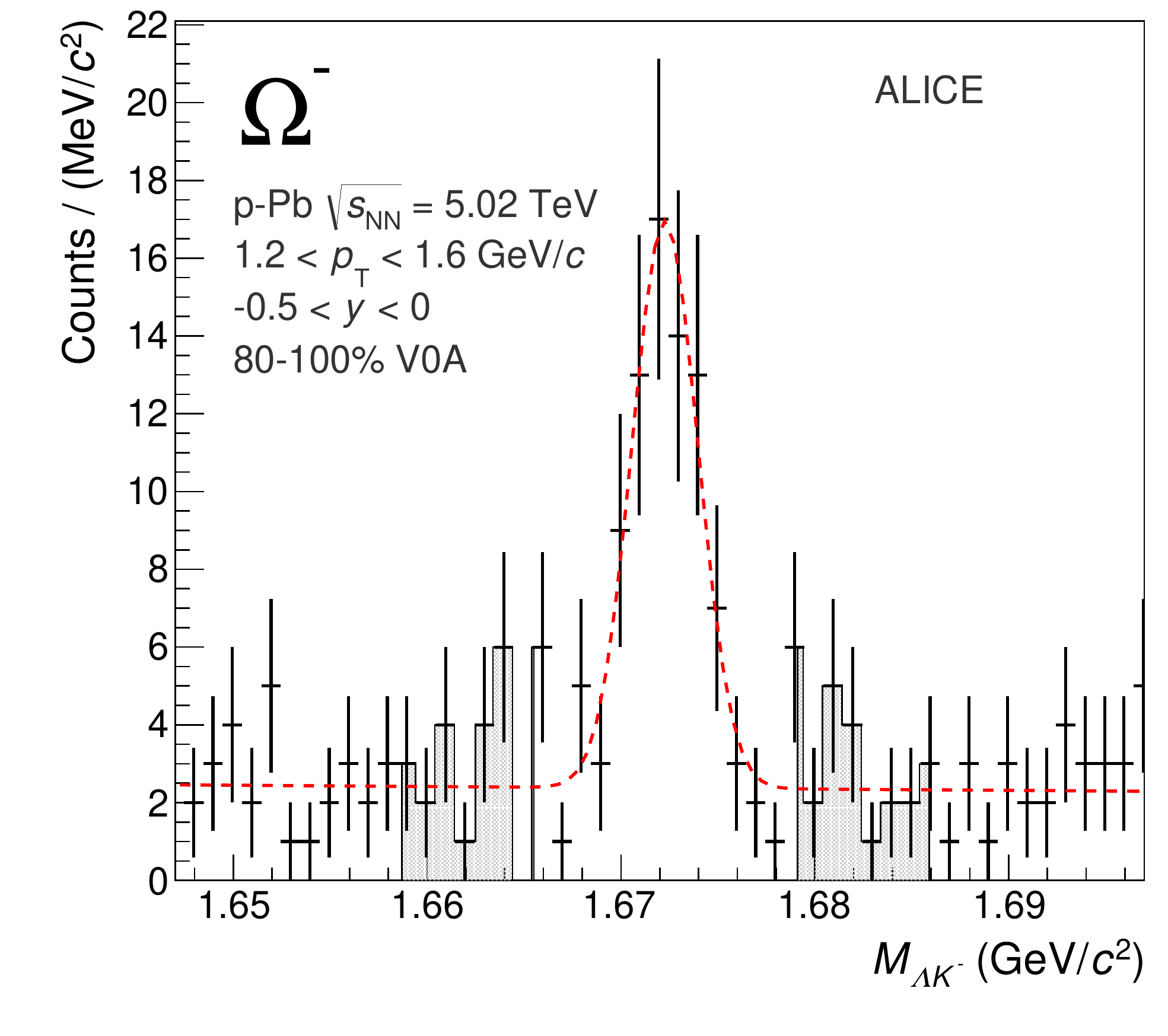} \\
    \end{flushleft}
    \caption{Invariant mass distributions of the $\Xi^{-}$ and $\Omega^{-}$
      in the 1.1-1.2 \GeVc~and 1.2-1.6 \GeVc~\pt\ bins
      respectively, fitted with a Gaussian peak and linear
      background (dashed red curves). The distributions for highest (left) and lowest
      (right) multiplicity classes are shown. The fits only serve to
      illustrate the peak position with respect to which the bands
      were defined and the linear background assumption for the applied
      signal extraction method.} 
    \label{fig:invmasses}
\end{figure*}

The invariant mass of the $\Xi$ and $\Omega$ hyperons was calculated
by assuming the known masses \cite{Agashe:2014kda} of the $\Lambda$ and of the bachelor
track. The mass was reconstructed twice for each cascade candidate,
once assuming the bachelor to be a $\pi$ and once a K. This allowed the
removal of an important fraction of the $\Omega$ background, which
contained a large contribution from the $\Xi$
candidates that pass the $\Omega$ selection criteria. Most of these false
$\Omega$ were removed discarding all candidates that could be
reconstructed as $\Xi$ with a mass within 10 \MeVmass~of the known mass \cite{Agashe:2014kda} of the
$\Xi$ baryon. Figure \ref{fig:invmasses} shows the invariant mass distributions for the
$\Xi^{-}$ and $\Omega^{-}$ hadrons in well populated \pt\ bins for the
lowest and highest multiplicity classes. 

For the signal extraction, a peak region was defined
within 4$\sigma$ of the mean of a Gaussian invariant mass peak for every
measured \pt\ interval. Adjacent background bands,
covering an equal combined mass interval as the peak region, were defined on both sides of that central 
region. This is illustrated in Figure \ref{fig:invmasses} with the
shaded bands on either side of the peak. The number of bin entries inside the
side-bands was subtracted from the number of candidates within
the peak region, assuming the background to be linear across the mass
range considered. 

The \pt ~distributions were corrected for detector acceptance and
reconstruction efficiencies. These were estimated with the use of
DPMJet \cite{DPMJet} simulated Monte Carlo (MC) events, which were propagated
through the detector with GEANT3 \cite{Brun:1994aa}. 

\subsection{Systematic uncertainties}

Systematic uncertainties due to the choice of selection
criteria were examined separately in each \pt\ interval of the measured
spectra. Individual settings were loosened and tightened, in order to
measure changes in the signal loss correction. For the $\Xi$ hyperons,
the signal extraction accounts for an uncertainty of around 2\% but reaches
5\% at low-\pt\ and in high multiplicity events, while for the $\Omega$, 
uncertainties of 3-5\% were measured. The uncertainty due to the topological
selections is around 2(3)\% for the main \pt\ region, and up to
3(5)\% at low momentum for $\Xi$($\Omega$). The constraint on the
\Vdecay mass window contributes to the total uncertainty with around
0.5(1)\%  and both the TPC tracking and identification cuts with 2(3)\%.
The proper decay length cut gives another 3(5)\% uncertainty at low
\pt. A 4\% error was added due to the material budget, and for the \Oms\ only, an
additional 3\% due to the mass hypothesis cut. All these
individual error contributions, which are listed in Table
\ref{tab:SystErros}, are added in quadrature. Apart from the low
momentum region, no \pt\ dependence is observed in the total
uncertainty. The total systematic error lies between 5-6(8)\% across
the whole spectrum, reaching up to 8(14)\% in the lowest \pt\ bins for the
$\Xi$($\Omega$) baryons.   

The fraction of the systematic error that is uncorrelated across
multiplicity was calculated by using the same method applied
in \cite{piKpLambda_pPb}, in which spectra deviations in specific
multiplicity classes were compared to those observed in the integrated
data sample. The choice of the topological
parameter values and the applied signal extraction method generates the
dominant contribution to the uncorrelated uncertainties across
multiplicity. These uncertainties were measured to be within 2\% in
the case of the $\Xi$ and 3\% in the case of the $\Omega$, which
constitutes a fraction that lies between 20 and 40$\%$ of the total
systematic uncertainties.

\begin{table*}
\begin{center}
\begin{tabular}{ l  l  l }
\hline
Source   & \Xis\ & \Oms\ \\
\hline       
Material budget                             & 4$\%$   & 4$\%$  \\
Competing mass hypothesis         &     -       & 3$\%$ \\
Topological variables                    &  2-3(5)\%     &   3-5\%   \\
Signal extraction                          &   2(5)\%         &  3(5)\%  \\           
Particle identification                   & 2$\%$         & 3$\%$ \\
Track selection                            & 2$\%$         & 3$\%$  \\
Proper decay length                     & 1(3)$\%$        &  2(5)$\%$  \\
\Vdecay mass window                  &   0.5\%       & 1\%   \\
  \\       

\end{tabular}
\end{center}
\caption{Contributions to the total systematic uncertainties for the
  \Xis\ and \Oms\ spectra measurements. The values in brackets indicate
  the maximum uncertainties measured for low \pt\ cascades (see text).}
\label{tab:SystErros}
\end{table*}

\section{Results}

\subsection{Transverse momentum spectra} 

\begin{figure*}[t!]
  \begin{flushleft}
    \includegraphics[width=0.495\textwidth]{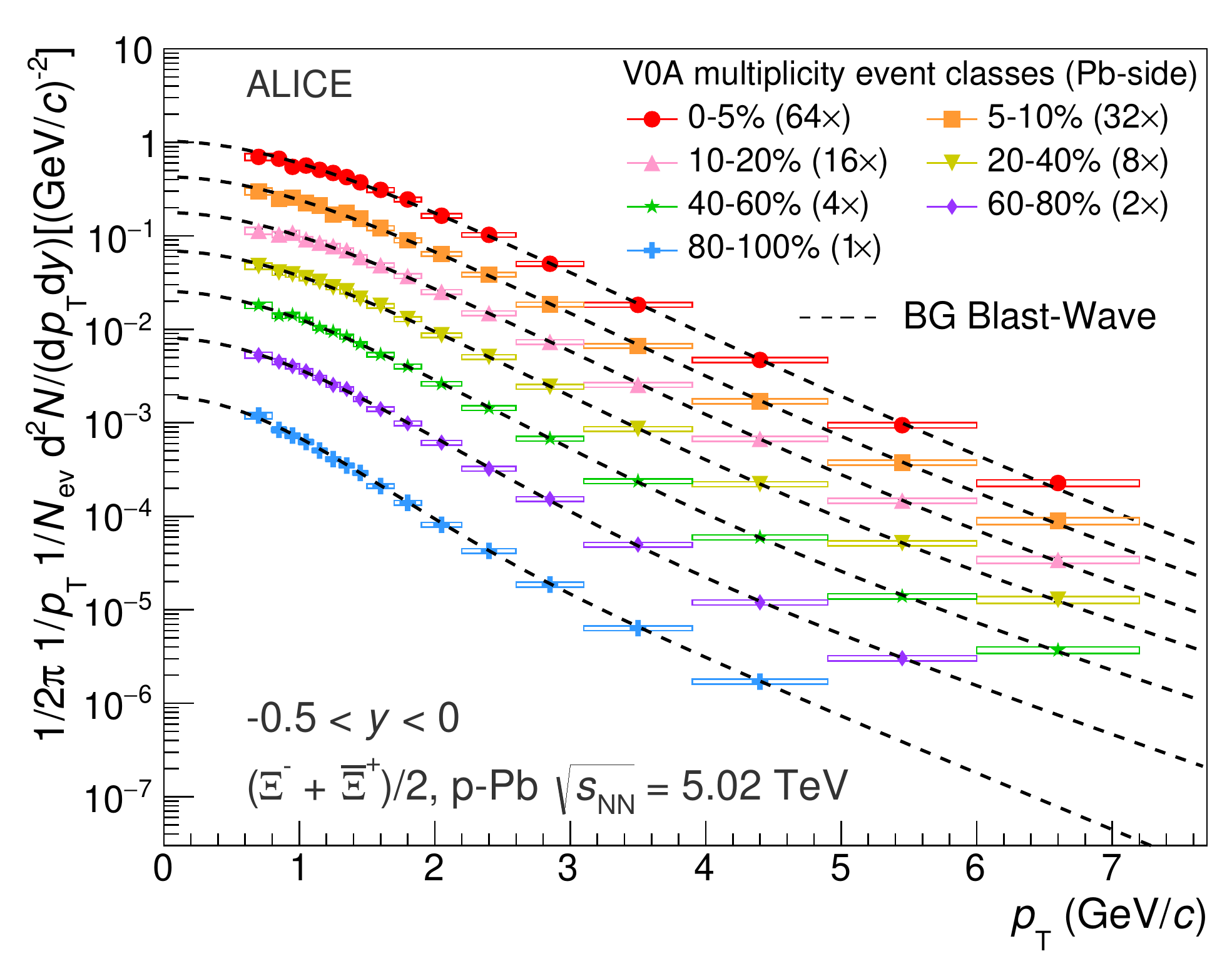}
    \includegraphics[width=0.495\textwidth]{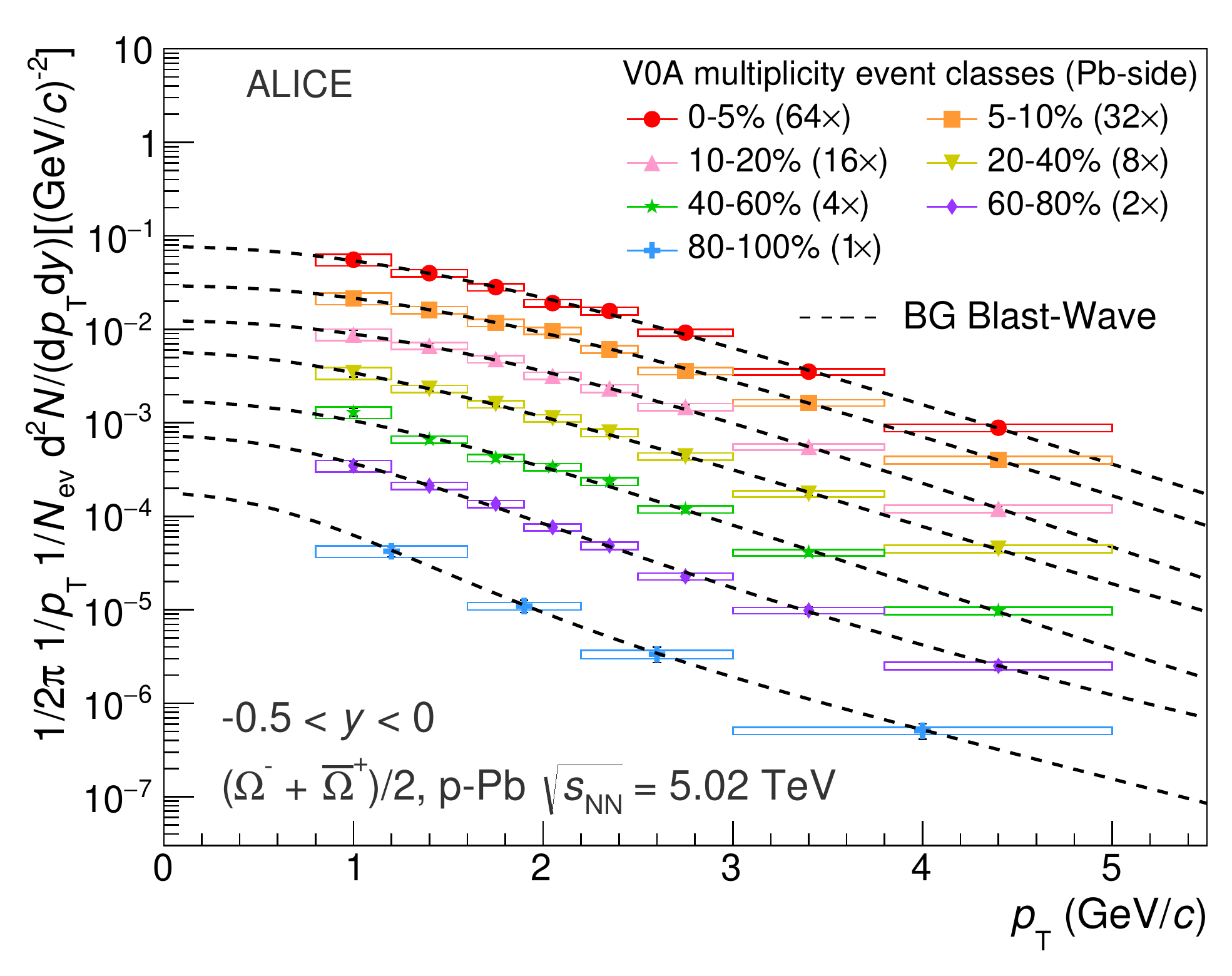}
    \end{flushleft}
    \caption{(colour online) Invariant \pt-differential yields of
      (\X+\Ix) and (\Om+\Mo) in different multiplicity classes. Data
      have been scaled by successive factors of 2 for better
      visibility. Statistical (bars), full systematic (boxes) and
      uncorrelated across multiplicity (transparent boxes)
      uncertainties are plotted. The dashed curves represent
      Blast-Wave fits to each individual distribution.} 
    \label{fig:spectra}
\end{figure*}  

The \pt\ distributions of \allpart\ in $-0.5 < y < 0$ are shown 
in Fig.~\ref{fig:spectra} for different multiplicity intervals, 
as defined in \cite{piKpLambda_pPb}. Since antiparticle and 
particle spectra are identical within uncertainties, the average of
the two is shown. The spectra exhibit a progressive flattening with
increasing multiplicity, which is qualitatively reminiscent of what is
observed in \PbPb\ \cite{PbPbmsALICE}. 

The calculation of \pt-integrated yields 
can be performed
by using data in the measured region and a
parametrisation-based extrapolation elsewhere. The Boltzmann-Gibbs
Blast-Wave (BG-BW) model  \cite{Schnedermann:1993ws} gives a good
description of each \pt\ spectrum and has been used as a tool for this
extrapolation. Other alternatives, such as the Levy-Tsallis \cite{Tsallis:1987eu} and Boltzmann distributions, were used for
estimating the systematic uncertainty due to the extrapolation. 

The extrapolation in the unmeasured \Xis\ (\Oms)  low-\pt\ region
grows progressively with decreasing multiplicity bins, from around
16$\%$(19$\%$) of the total yield in the 0--5$\%$ multiplicity class to around
27$\%$(40$\%$) in the 80--100$\%$ class. The systematic
uncertainty assigned to the yield due to the extrapolation technique is  
2.8$\%$(7.8$\%$) for high multiplicities and rises to
5.2$\%$(14.5$\%$) in the case where the fraction of the extrapolated
yield is highest.   

\if\includeBW1
\subsection{Comparison to Blast-Wave model} 

In order to investigate whether the observed spectral shapes are consistent with a system
that exhibits hydrodynamical radial expansion, the measured distributions have been 
further studied in the context of the BG-BW model \cite{Schnedermann:1993ws}. 
This model assumes a locally thermalised medium that expands collectively with 
a common velocity field and then undergoes an instantaneous freeze-out. 
In this framework, a simultaneous fit to identified particle spectra allows for 
the determination of common freeze-out parameters. These can be used to predict the \pt\ distribution 
for other particle species in a collective expansion picture. It should be noted that such a simultaneous 
fit differs from the individual fits mentioned in the previous section
and used only for extrapolating the spectra.

The \allpart\ \pt\ spectra in the  0--5$\%$ and 80--100$\%$
multiplicity classes are compared to predictions  from the BG-BW model
with parameters acquired  from a simultaneous fit to $\pi^{\pm}$,
K$^{\pm}$, p(\pbar) and $\Lambda$($\bar{\Lambda}$)
\cite{piKpLambda_pPb} in Fig.~\ref{fig:BGBW}. The model describes the
measured shapes within  uncertainties up to a \pt\ of approximately
4~\GeVc~for $\Xi$ and 5~\GeVc~for $\Omega$ in the highest multiplicity
class. This indicates that multi-strange hadrons also follow a common
motion with the lighter hadrons and is suggestive of the presence of
radial flow in p--Pb collisions. However, it is worth noting that
some final state effects could also modify the spectra in a similar
manner to radial flow. For example, PYTHIA \cite{Pythia6point4}
implements the colour reconnection mechanism, which fuses strings
originating from independent parton interactions, leading to fewer but
more energetic hadrons, which has been shown to mimic radial flow
\cite{PhysRevLett.111.042001}.   

Applying the same technique to results from the lower multiplicity
classes reveals that the agreement of the data with the Blast-Wave
predictions become progressively worse. The comparison between lowest
and highest multiplicity cases can be seen in Fig.~\ref{fig:BGBW},
where their respective ratios to the model predictions are shown in
the lower panels. These observations indicate that common kinetic
freeze-out conditions are able to better describe the spectra in high
multiplicity \pPb~collisions. 

The multi-strange baryon spectra in central \PbPb~collisions
\cite{PbPbmsALICE} have also been investigated in a common
freeze-out scenario \cite{PhysRevC.90.054912,Melo:2015iva} and similar
studies were performed for \AuAu\ collisions
\cite{Adams2005102}. In contrast to high multiplicity \pPb~collisions,
where all stable and long-lived hadron spectra are compatible with a
single set of kinetic freeze-out conditions (the temperature
T$_\mathrm{fo}$ and the mean transverse flow velocity \avbT),
multi-strange particles in central heavy-ion collisions seem to
experience less transverse flow and may freeze out earlier in the
evolution of the system when compared to most of the other hadrons. 

\begin{figure*}[t!]
  \begin{flushleft}
  \includegraphics[width=0.495\textwidth]{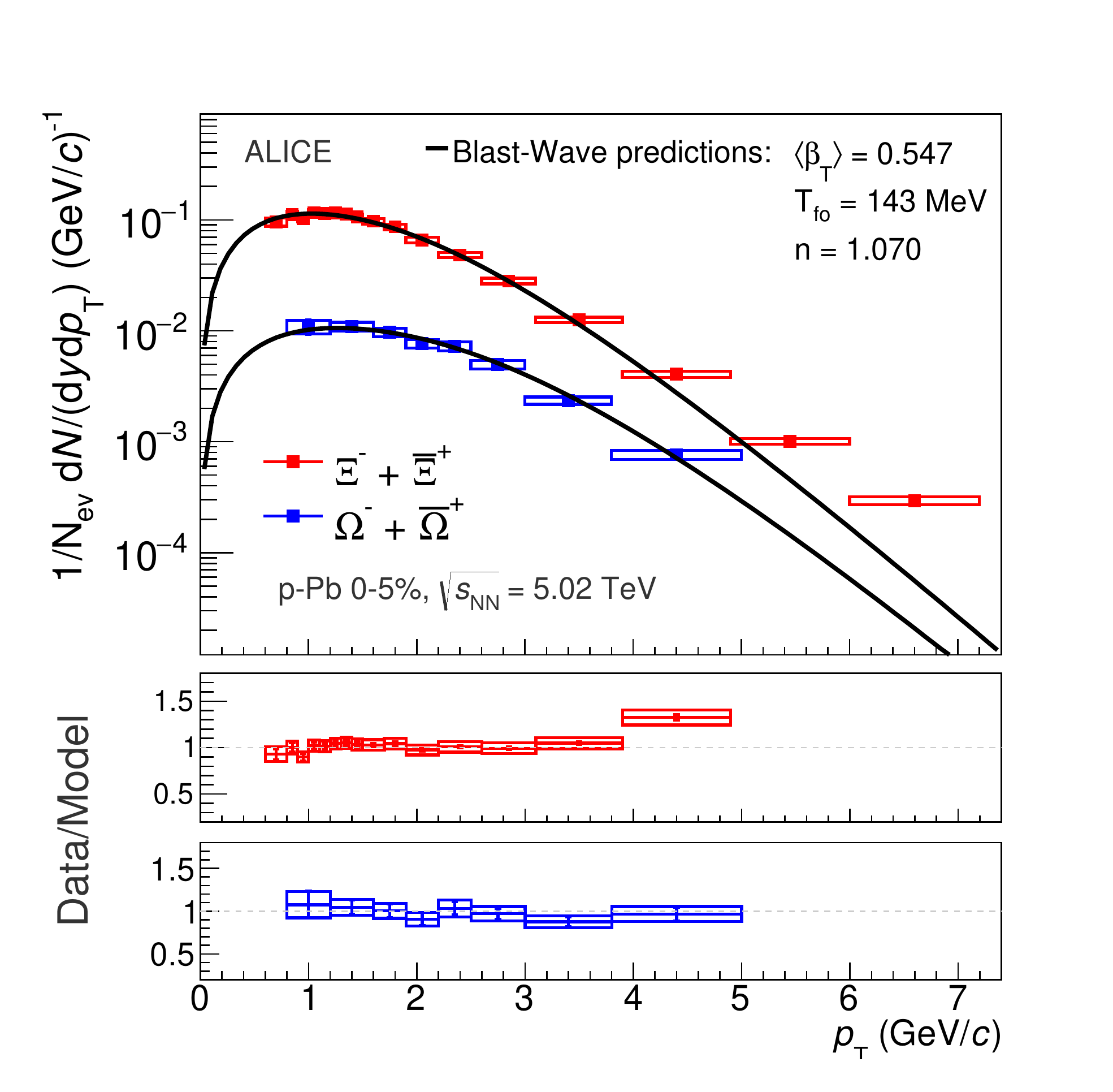}
  \includegraphics[width=0.495\textwidth]{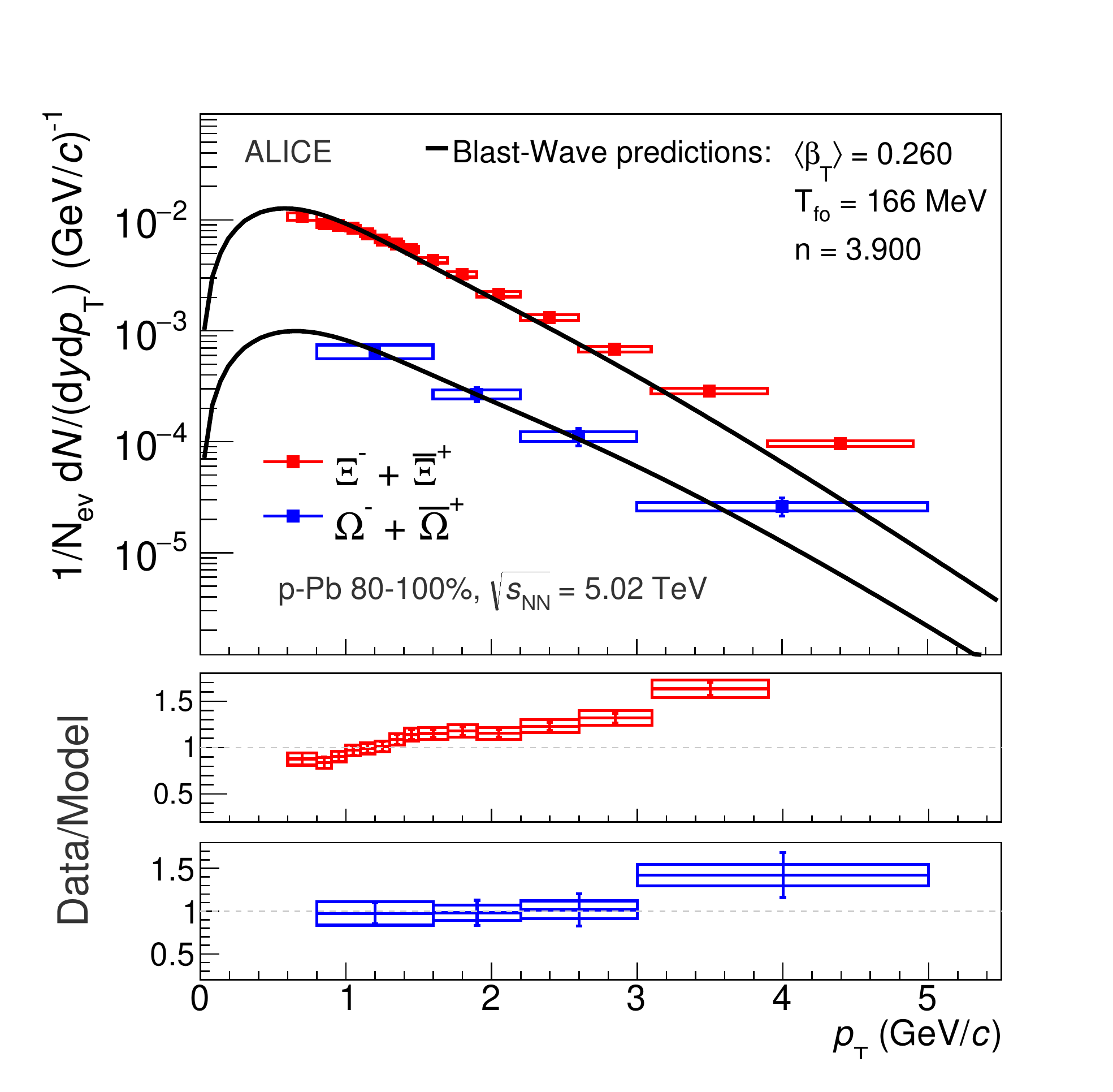}
  \end{flushleft}
  \caption{(colour online) (\X+\Ix) and (\Om+\Mo) \pt\ spectra
    in the 0--5$\%$ (left) and 80--100$\%$ (right) multiplicity classes compared
    to predictions from the BG-BW model (upper panels) with the ratios
    on a linear scale (lower panels). The parameters are based on
    simultaneous fits to lighter hadrons \cite{piKpLambda_pPb}.  See
    text for details. } 
  \label{fig:BGBW}
\end{figure*}
\fi

\subsection{Hyperon to pion ratios} 

\begin{table}[t] 
  \begin{center}
    \begin{tabular*}{\linewidth}{@{\extracolsep{\fill}}l  l  l  l }
      &&& \\[-0.7em]
      \hline
      Event class & $\avg{\dNdeta}$                           & dN/dy(\X + \Ix) & dN/dy(\Om + \Mo) \\[-0.2em]
      & \footnotesize{$|\hlab|<0.5$}       &                &   \\ [+0.1em]
      \hline
      &&& \\[-0.4em]
      0--5\%    &    45   $\pm$ 1      & 0.2354$\pm$0.0020$\pm$0.0161 & 0.0260$\pm$0.0011$\pm$0.0034 \\[0.3em]
      5--10\%   &   36.2 $\pm$ 0.8  & 0.1861$\pm$0.0016$\pm$0.0138 & 0.0215$\pm$0.0008$\pm$0.0029 \\[0.3em]
      10--20\%  &  30.5 $\pm$ 0.7  &0.1500$\pm$0.0010$\pm$0.0112 & 0.0167$\pm$0.0006$\pm$0.0022 \\[0.3em]
      20--40\%  &  23.2 $\pm$ 0.5  &0.1100$\pm$0.0006$\pm$0.0085 & 0.0120$\pm$0.0005$\pm$0.0016 \\[0.3em] 
      40--60\%  &  16.1 $\pm$ 0.4  &0.0726$\pm$0.0006$\pm$0.0065 & 0.0072$\pm$0.0003$\pm$0.0010 \\[0.3em]
      60--80\%  &  9.8  $\pm$ 0.2   &0.0398$\pm$0.0004$\pm$0.0031 &0.0042$\pm$0.0002$\pm$0.0006\\[0.3em]
      80--100\% & 4.3  $\pm$ 0.1   &0.0143$\pm$0.0003$\pm$0.0015 &0.0013$\pm$0.0003$\pm$0.0003\\[0.3em]
    \end{tabular*}
    \caption{The mid-rapidity $\avg{\dNdeta}$ values for each of the 7
      multiplicity classes and the \X + \Ix and \Om + \Mo integrated
      yields per unit rapidity normalised to the visible cross
      section. The statistical uncertainty on the yields is followed
      by the sytematic uncertainty.}
    \label{tab:yields}
  \end{center}
\end{table}

\begin{figure*}[t!]
  \begin{flushleft}
  \includegraphics[width=0.495\textwidth]{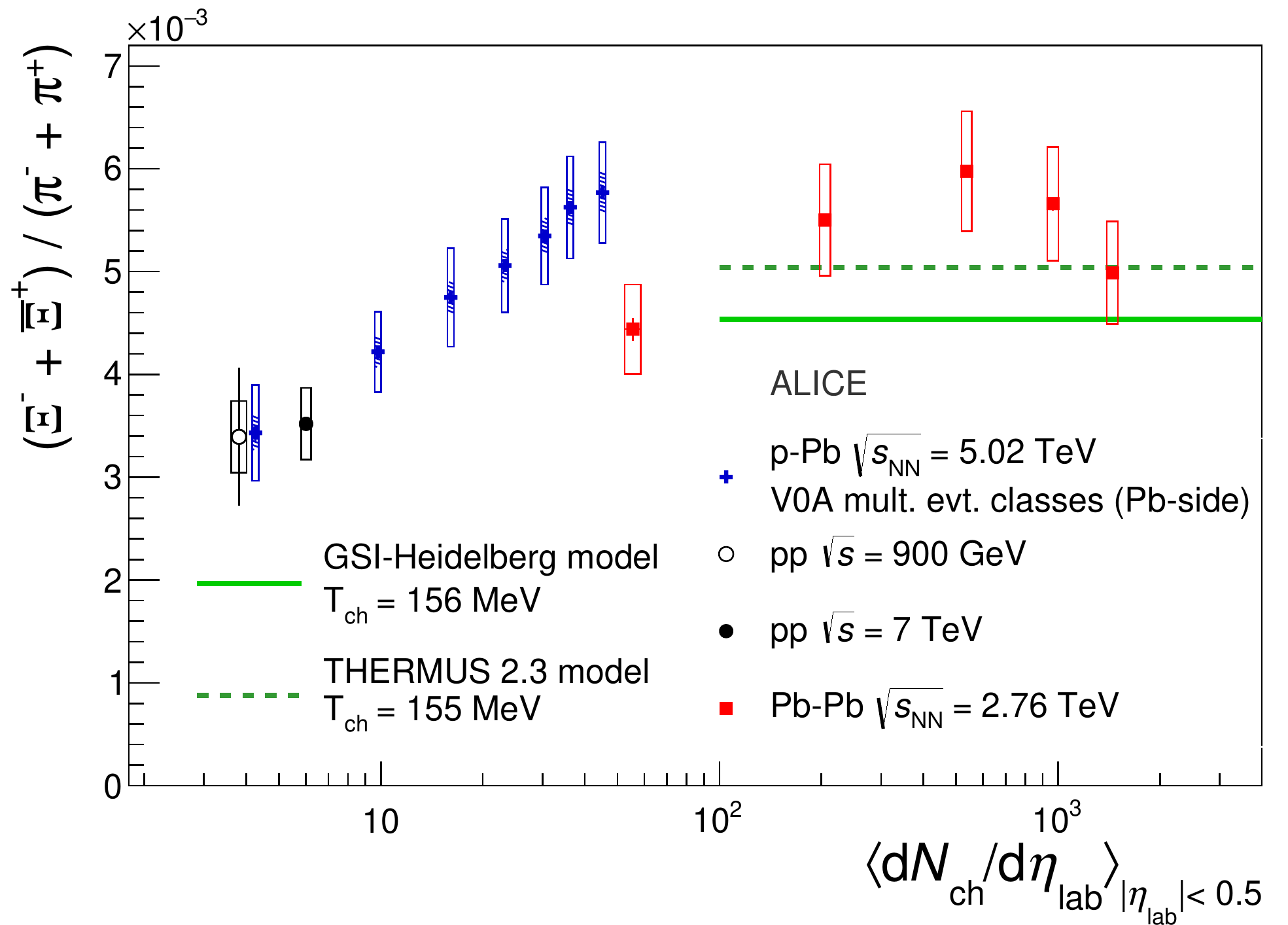}
  \includegraphics[width=0.495\textwidth]{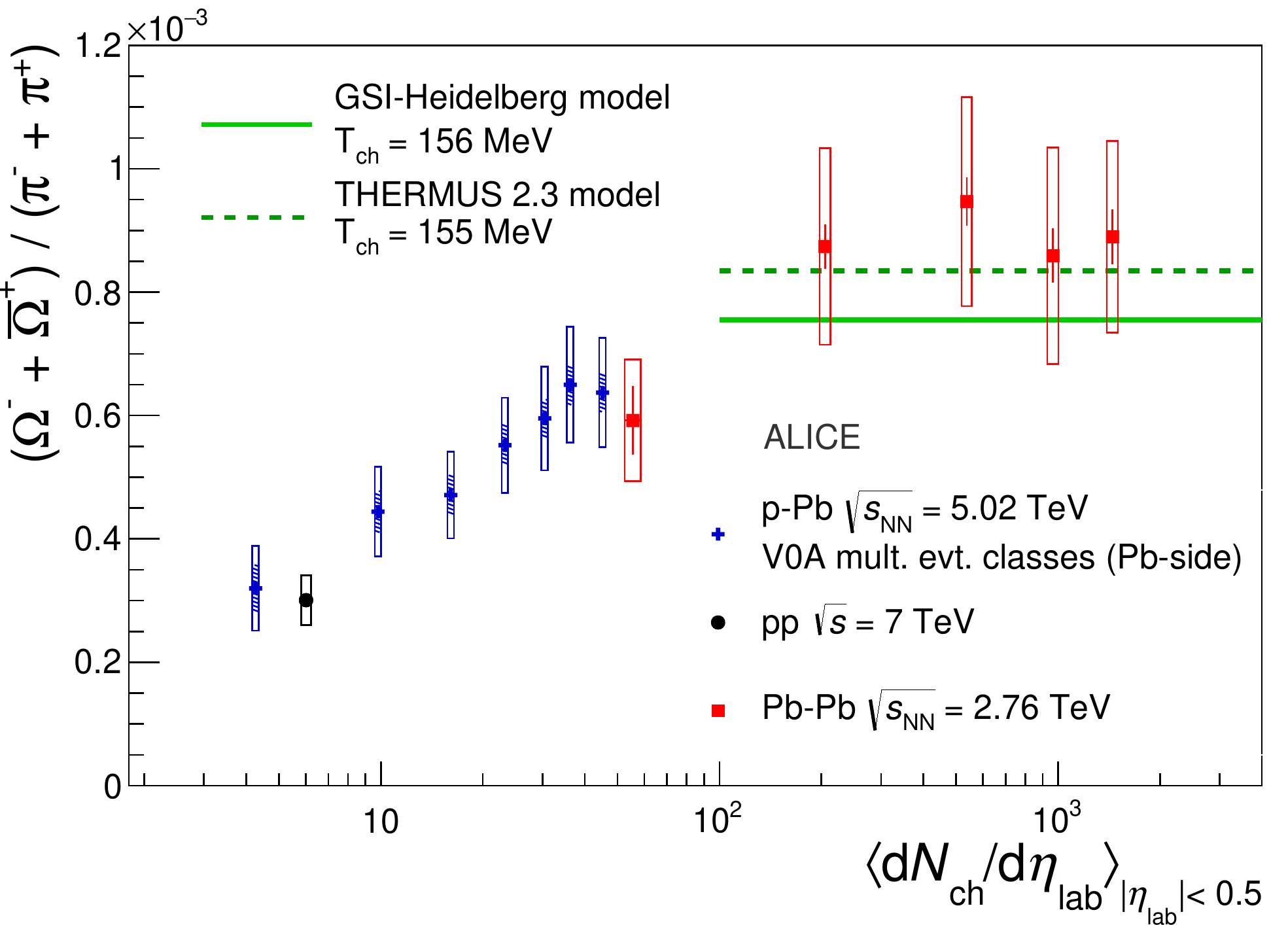}
  \end{flushleft}
  \caption{(colour online) (\X+\Ix)/(\pip+\pim) (left) and
    (\Om+\Mo)/(\pip+\pim) (right) ratios as a function of
    $\avg{\dNdeta}$ for all three colliding systems. The ratios for the seven multiplicity
    classes in \pPb\ data lie between the Minimum Bias pp (\s~=~900 GeV
    \cite{ppSpectra900ALICE,Strangenesspp900ALICE} and \s~=~7 TeV
    \cite{ms7ppALICE,ppSpectra7TeVALICE}) and peripheral \PbPb\
    results. The \PbPb\ points \cite{PbPbmsALICE} represent, from left
    to right, the 60-80$\%$, 40-60$\%$, 20-40$\%$ and 10-20$\%$ and
    0-10$\%$  centrality classes. The chemical equilibrium predictions
    by the GSI-Heidelberg \cite{Andronic:2009ht} and the \thermus~2.3
    \cite{Wheaton200984} models are represented by the horizontal
    lines. }
  \label{fig:CascadeOverPiRatios}
\end{figure*}

The measured integrated yields in the seven multiplicity classes are given in
Table \ref{tab:yields}. To study the relative production
of strangeness and compare it with results in pp and \PbPb\
collisions, the yield ratios to pions were calculated as a function of
charged particle multiplicity. Both the (\X+\Ix)/(\pip+\pim) and (\Om+\Mo)/(\pip+\pim)
ratios are observed to increase as a function of multiplicity, as seen in
Fig.~\ref{fig:CascadeOverPiRatios}. The relative increase is more
pronounced for the \Om\ and \Mo\ than for \X\ and \Ix, being
approximately 100$\%$ for the former and 60$\%$ for the latter.
These relative increases are larger than the 30$\%$ increase observed
for the $\Lambda/\pi$ ratio  \cite{piKpLambda_pPb}, indicating that
strangeness content may control the rate of increase with
multiplicity.  

These ratios are further compared to measurements performed in the pp
\cite{ms7ppALICE,ppSpectra7TeVALICE} and \PbPb\ \cite{PbPbmsALICE}
collision systems. The (\X+\Ix)/(\pip+\pim) ratio for the highest \pPb\
multiplicity is compatible with the \PbPb\ measurements in
the Pb--Pb 0-60\% centrality range and the (\Om+\Mo)/(\pip+\pim) reaches a
value slightly below its \PbPb\ equivalent in this centrality range,
although the error bars still overlap. It is also
noteworthy that the values obtained for the \pPb~80--100$\%$
multiplicity event class are similar to the ones measured in minimum
bias pp collisions.   

Finally, the hyperon to pion ratios can also be compared with the
values in the Grand  Canonical (GC) limit obtained from global fits to
\PbPb\ data. Two different implementations of the thermal model are
shown  in Fig.~\ref{fig:CascadeOverPiRatios}, where the dashed lines
represent the values from the \thermus~2.3 model \cite{Wheaton200984}
and the solid lines represent predictions from the GSI-Heidelberg
model \cite{Andronic:2009ht}. Both models provide values that are
consistent with the most central \PbPb\ measurements.

\begin{figure}[t]
  \centering
  \includegraphics[width=0.7\textwidth]{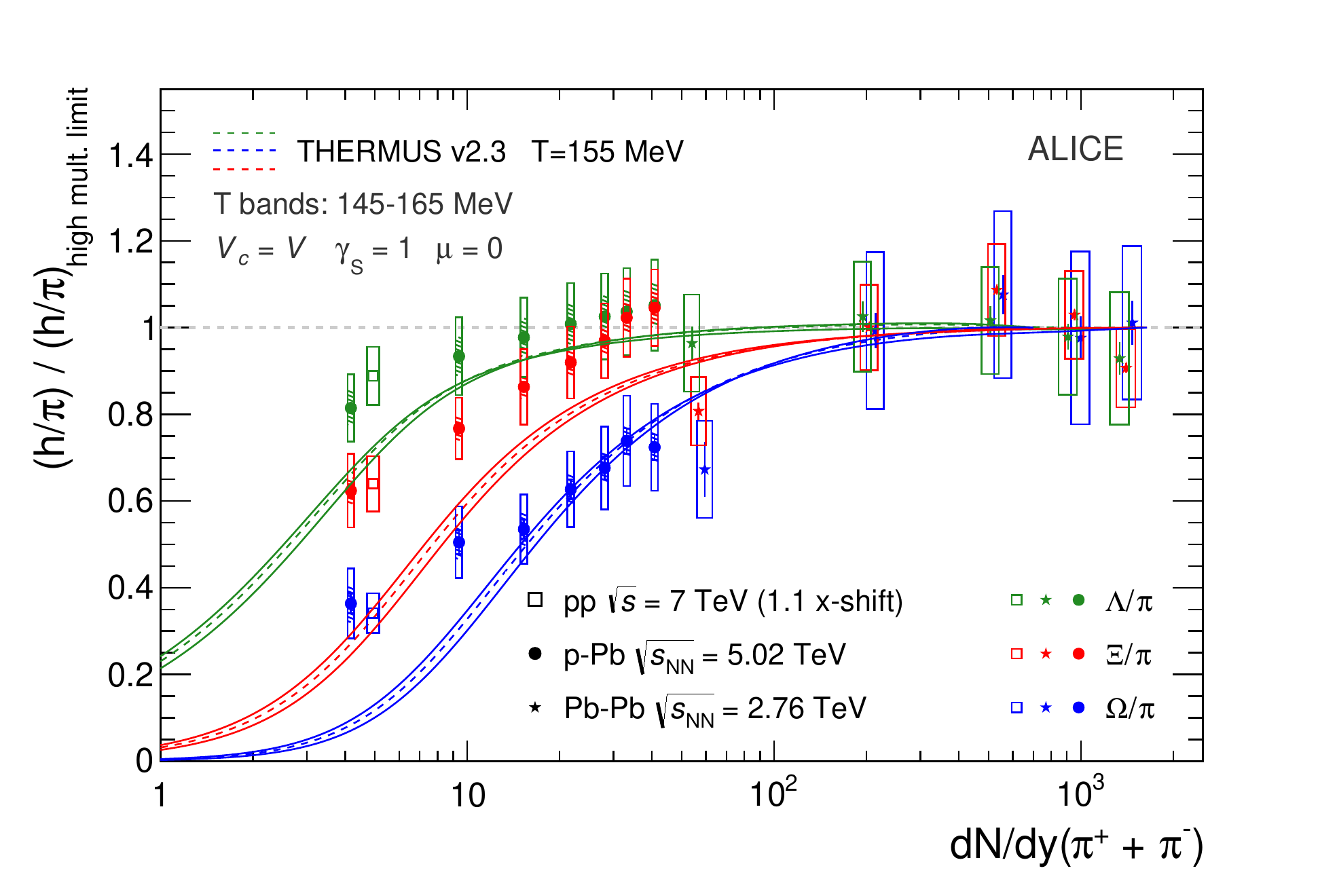}  
  \caption{(colour online) Hyperon to pion ratios as a function of
    pion yields for pp, \pPb\ and \PbPb\ colliding systems compared to
    the \thermus\ \cite{Wheaton200984} strangeness suppression model prediction, in which
    only the system size is varied. The h/$\pi$ are the ratios of the particle and
    antiparticle sums, except for the 2$\Lambda$/($\pi^{-}+\pi^{+}$)
    data points in pp \cite{Strangenesspp900ALICE},
    \pPb~\cite{piKpLambda_pPb} and \PbPb~\cite{KosLambdaPbPbALICe}. All
    values are normalised to the high multiplicity limit, which is
    given by the mean of the 0-60\% highest multiplicity
    \PbPb~measurements for the data and by the GC limit for the model.}
  \label{fig:CanonicalSuppressiontest}
\end{figure}

In small multiplicity environments such as those produced in \pPb\
collisions, a grand canonical statistical description may not be
appropriate. Instead, local conservation laws might  play an important
role. The evolution of hyperon to pion ratios in terms of the event
multiplicity can be calculated with a Strangeness Canonical (SC) model
implemented in \thermus~\cite{Wheaton200984}. This model applies a local conservation
law to the strangeness quantum number within a correlation volume $V_{c}$
while treating the baryon and charge quantum numbers grand-canonically
within the fireball volume $V$. This implies a decrease of the strangeness yields
with respect to the pion yields with a shrinking system size. To model this canonical suppression
effect as a function of pion rapidity density, yield calculations were
repeated for varying system sizes. Strangeness conservation was
imposed within the size of the fireball ($V_{c}=V$), and the
strangeness saturation parameter $\gamma_{\rm{S}}$  was fixed to 1,
thus changes in the hadron to pion ratios were due to the variations of the restraints on
the system size only. The chemical potentials ($\mu$) of the conserved
strangeness, baryon and electric charge quantum numbers were set to
zero. The obtained suppression curves for $\Lambda$, $\Xi$ and
$\Omega$ are shown in Fig.~\ref{fig:CanonicalSuppressiontest} for a temperature of 155 MeV,
the value extracted from a GC global fit to high multiplicity \PbPb\
data, with a variation of $\pm$10 MeV (solid lines). Both the data and
model points were normalised to the high multiplicity limit. For the
data, this limit is the mean hyperon to pion ratio in the 0-60$\%$
most central \PbPb\ events, whereas for the model it corresponds to
the GC limit. The theoretical curves for strangeness suppression
computed with \thermus~are in qualitative agreement with the effect
observed in the data.  

\section{Conclusions}

In summary, a measurement of the \pt\ spectra of \allpart\ for seven multiplicity classes in
\pPb\ collisions at \snn=~5.02~TeV at the LHC has been presented. These measurements represent an important contribution to 
the understanding of strangeness production, as hyperon production rates are now measured at LHC
energies over a large range in charged--particle multiplicity, from pp to central \PbPb\ collisions. 

\if\includeBW1
The multi-strange baryon spectra exhibit a progressive flattening with
increasing multiplicity suggesting the presence of radial flow. A
comparison with the Boltzmann-Gibbs Blast-Wave model indicates a
common kinetic freeze-out with lighter hadrons in the highest
multiplicity \pPb\ collisions. This is in contrast to higher
multiplicity heavy-ion collisions where there is an indication for an
earlier freeze-out of these particles. 
\fi

For the first time, the lifting of strangeness suppression with system
size has been observed with measurements in a single collision
system. Hyperon to pion ratios are shown to increase with multiplicity
in \pPb\ collisions from the values measured in pp to those observed
in \PbPb. The rate of increase is more pronounced  for particles with
higher strangeness content. Comparing these results to the trends
observed in statistical hadronisation models that conserve strangeness
across the created system indicates that the behaviour is
qualitatively consistent with the lifting of canonical suppression
with increasing multiplicity.  



%% file: acknowledgements.tex

The ALICE Collaboration would like to thank all its engineers and technicians for their invaluable contributions to the construction of the experiment and the CERN accelerator teams for the outstanding performance of the LHC complex.
The ALICE Collaboration gratefully acknowledges the resources and support provided by all Grid centres and the Worldwide LHC Computing Grid (WLCG) collaboration.
The ALICE Collaboration acknowledges the following funding agencies for their support in building and
running the ALICE detector:
State Committee of Science,  World Federation of Scientists (WFS)
and Swiss Fonds Kidagan, Armenia;
Conselho Nacional de Desenvolvimento Cient\'{\i}fico e Tecnol\'{o}gico (CNPq), Financiadora de Estudos e Projetos (FINEP),
Funda\c{c}\~{a}o de Amparo \`{a} Pesquisa do Estado de S\~{a}o Paulo (FAPESP);
National Natural Science Foundation of China (NSFC), the Chinese Ministry of Education (CMOE)
and the Ministry of Science and Technology of China (MSTC);
Ministry of Education and Youth of the Czech Republic;
Danish Natural Science Research Council, the Carlsberg Foundation and the Danish National Research Foundation;
The European Research Council under the European Community's Seventh Framework Programme;
Helsinki Institute of Physics and the Academy of Finland;
French CNRS-IN2P3, the `Region Pays de Loire', `Region Alsace', `Region Auvergne' and CEA, France;
German Bundesministerium fur Bildung, Wissenschaft, Forschung und Technologie (BMBF) and the Helmholtz Association;
General Secretariat for Research and Technology, Ministry of Development, Greece;
National Research, Development and Innovation Office (NKFIH), Hungary;
Department of Atomic Energy and Department of Science and Technology of the Government of India;
Istituto Nazionale di Fisica Nucleare (INFN) and Centro Fermi -
Museo Storico della Fisica e Centro Studi e Ricerche ``Enrico Fermi'', Italy;
Japan Society for the Promotion of Science (JSPS) KAKENHI and MEXT, Japan;
Joint Institute for Nuclear Research, Dubna;
National Research Foundation of Korea (NRF);
Consejo Nacional de Cienca y Tecnologia (CONACYT), Direccion General de Asuntos del Personal Academico(DGAPA), M\'{e}xico, Amerique Latine Formation academique - 
European Commission~(ALFA-EC) and the EPLANET Program~(European Particle Physics Latin American Network);
Stichting voor Fundamenteel Onderzoek der Materie (FOM) and the Nederlandse Organisatie voor Wetenschappelijk Onderzoek (NWO), Netherlands;
Research Council of Norway (NFR);
National Science Centre, Poland;
Ministry of National Education/Institute for Atomic Physics and National Council of Scientific Research in Higher Education~(CNCSI-UEFISCDI), Romania;
Ministry of Education and Science of Russian Federation, Russian
Academy of Sciences, Russian Federal Agency of Atomic Energy,
Russian Federal Agency for Science and Innovations and The Russian
Foundation for Basic Research;
Ministry of Education of Slovakia;
Department of Science and Technology, South Africa;
Centro de Investigaciones Energeticas, Medioambientales y Tecnologicas (CIEMAT), E-Infrastructure shared between Europe and Latin America (EELA), 
Ministerio de Econom\'{i}a y Competitividad (MINECO) of Spain, Xunta de Galicia (Conseller\'{\i}a de Educaci\'{o}n),
Centro de Aplicaciones Tecnológicas y Desarrollo Nuclear (CEA\-DEN), Cubaenerg\'{\i}a, Cuba, and IAEA (International Atomic Energy Agency);
Swedish Research Council (VR) and Knut $\&$ Alice Wallenberg
Foundation (KAW);
Ukraine Ministry of Education and Science;
United Kingdom Science and Technology Facilities Council (STFC);
The United States Department of Energy, the United States National
Science Foundation, the State of Texas, and the State of Ohio;
Ministry of Science, Education and Sports of Croatia and  Unity through Knowledge Fund, Croatia;
Council of Scientific and Industrial Research (CSIR), New Delhi, India;
Pontificia Universidad Cat\'{o}lica del Per\'{u}.

%% file: Alice_Authorlist_2015-Nov-27.tex


\begingroup
\small
\begin{flushleft}
J.~Adam\Irefn{org40}\And
D.~Adamov\'{a}\Irefn{org84}\And
M.M.~Aggarwal\Irefn{org88}\And
G.~Aglieri Rinella\Irefn{org36}\And
M.~Agnello\Irefn{org110}\And
N.~Agrawal\Irefn{org48}\And
Z.~Ahammed\Irefn{org132}\And
S.~Ahmad\Irefn{org19}\And
S.U.~Ahn\Irefn{org68}\And
S.~Aiola\Irefn{org136}
% ...
\renewcommand\labelenumi{\textsuperscript{\theenumi}~}

\section*{Affiliation notes}
\renewcommand\theenumi{\roman{enumi}}
\begin{Authlist}
\item \Adef{0}Deceased
\item \Adef{idp1767392}{Also at: Georgia State University, Atlanta, Georgia, United States}
\item \Adef{idp3128880}{Also at: Also at Department of Applied Physics, Aligarh Muslim University, Aligarh, India}
\item \Adef{idp3835440}{Also at: M.V. Lomonosov Moscow State University, D.V. Skobeltsyn Institute of Nuclear, Physics, Moscow, Russia}
\end{Authlist}

\section*{Collaboration Institutes}
\renewcommand\theenumi{\arabic{enumi}~}
\begin{Authlist}

\item \Idef{org1}A.I. Alikhanyan National Science Laboratory (Yerevan Physics Institute) Foundation, Yerevan, Armenia
\item \Idef{org2}Benem\'{e}rita Universidad Aut\'{o}noma de Puebla, Puebla, Mexico
\item \Idef{org3}Bogolyubov Institute for Theoretical Physics, Kiev, Ukraine
\item \Idef{org4}Bose Institute, Department of Physics and Centre for Astroparticle Physics and Space Science (CAPSS), Kolkata, India
\item \Idef{org5}Budker Institute for Nuclear Physics, Novosibirsk, Russia
\item \Idef{org6}California Polytechnic State University, San Luis Obispo, California, United States
\item \Idef{org7}Central China Normal University, Wuhan, China
\item \Idef{org8}Centre de Calcul de l'IN2P3, Villeurbanne, France
\item \Idef{org9}Centro de Aplicaciones Tecnol\'{o}gicas y Desarrollo Nuclear (CEADEN), Havana, Cuba
\item \Idef{org10}Centro de Investigaciones Energ\'{e}ticas Medioambientales y Tecnol\'{o}gicas (CIEMAT), Madrid, Spain
\item \Idef{org11}Centro de Investigaci\'{o}n y de Estudios Avanzados (CINVESTAV), Mexico City and M\'{e}rida, Mexico
\item \Idef{org12}Centro Fermi - Museo Storico della Fisica e Centro Studi e Ricerche ``Enrico Fermi'', Rome, Italy
\item \Idef{org13}Chicago State University, Chicago, Illinois, USA
\item \Idef{org14}China Institute of Atomic Energy, Beijing, China
\item \Idef{org15}Commissariat \`{a} l'Energie Atomique, IRFU, Saclay, France
\item \Idef{org16}COMSATS Institute of Information Technology (CIIT), Islamabad, Pakistan
\item \Idef{org17}Departamento de F\'{\i}sica de Part\'{\i}culas and IGFAE, Universidad de Santiago de Compostela, Santiago de Compostela, Spain
\item \Idef{org18}Department of Physics and Technology, University of Bergen, Bergen, Norway
\item \Idef{org19}Department of Physics, Aligarh Muslim University, Aligarh, India
\item \Idef{org20}Department of Physics, Ohio State University, Columbus, Ohio, United States
\item \Idef{org21}Department of Physics, Sejong University, Seoul, South Korea
\item \Idef{org22}Department of Physics, University of Oslo, Oslo, Norway
\item \Idef{org23}Dipartimento di Elettrotecnica ed Elettronica del Politecnico, Bari, Italy
\item \Idef{org24}Dipartimento di Fisica dell'Universit\`{a} 'La Sapienza' and Sezione INFN Rome, Italy
\item \Idef{org25}Dipartimento di Fisica dell'Universit\`{a} and Sezione INFN, Cagliari, Italy
\item \Idef{org26}Dipartimento di Fisica dell'Universit\`{a} and Sezione INFN, Trieste, Italy
\item \Idef{org27}Dipartimento di Fisica dell'Universit\`{a} and Sezione INFN, Turin, Italy
\item \Idef{org28}Dipartimento di Fisica e Astronomia dell'Universit\`{a} and Sezione INFN, Bologna, Italy
\item \Idef{org29}Dipartimento di Fisica e Astronomia dell'Universit\`{a} and Sezione INFN, Catania, Italy
\item \Idef{org30}Dipartimento di Fisica e Astronomia dell'Universit\`{a} and Sezione INFN, Padova, Italy
\item \Idef{org31}Dipartimento di Fisica `E.R.~Caianiello' dell'Universit\`{a} and Gruppo Collegato INFN, Salerno, Italy
\item \Idef{org32}Dipartimento di Scienze e Innovazione Tecnologica dell'Universit\`{a} del  Piemonte Orientale and Gruppo Collegato INFN, Alessandria, Italy
\item \Idef{org33}Dipartimento Interateneo di Fisica `M.~Merlin' and Sezione INFN, Bari, Italy
\item \Idef{org34}Division of Experimental High Energy Physics, University of Lund, Lund, Sweden
\item \Idef{org35}Eberhard Karls Universit\"{a}t T\"{u}bingen, T\"{u}bingen, Germany
\item \Idef{org36}European Organization for Nuclear Research (CERN), Geneva, Switzerland
\item \Idef{org37}Excellence Cluster Universe, Technische Universit\"{a}t M\"{u}nchen, Munich, Germany
\item \Idef{org38}Faculty of Engineering, Bergen University College, Bergen, Norway
\item \Idef{org39}Faculty of Mathematics, Physics and Informatics, Comenius University, Bratislava, Slovakia
\item \Idef{org40}Faculty of Nuclear Sciences and Physical Engineering, Czech Technical University in Prague, Prague, Czech Republic
\item \Idef{org41}Faculty of Science, P.J.~\v{S}af\'{a}rik University, Ko\v{s}ice, Slovakia
\item \Idef{org42}Faculty of Technology, Buskerud and Vestfold University College, Vestfold, Norway
\item \Idef{org43}Frankfurt Institute for Advanced Studies, Johann Wolfgang Goethe-Universit\"{a}t Frankfurt, Frankfurt, Germany
\item \Idef{org44}Gangneung-Wonju National University, Gangneung, South Korea
\item \Idef{org45}Gauhati University, Department of Physics, Guwahati, India
\item \Idef{org46}Helsinki Institute of Physics (HIP), Helsinki, Finland
\item \Idef{org47}Hiroshima University, Hiroshima, Japan
\item \Idef{org48}Indian Institute of Technology Bombay (IIT), Mumbai, India
\item \Idef{org49}Indian Institute of Technology Indore, Indore (IITI), India
\item \Idef{org50}Inha University, Incheon, South Korea
\item \Idef{org51}Institut de Physique Nucl\'eaire d'Orsay (IPNO), Universit\'e Paris-Sud, CNRS-IN2P3, Orsay, France
\item \Idef{org52}Institut f\"{u}r Informatik, Johann Wolfgang Goethe-Universit\"{a}t Frankfurt, Frankfurt, Germany
\item \Idef{org53}Institut f\"{u}r Kernphysik, Johann Wolfgang Goethe-Universit\"{a}t Frankfurt, Frankfurt, Germany
\item \Idef{org54}Institut f\"{u}r Kernphysik, Westf\"{a}lische Wilhelms-Universit\"{a}t M\"{u}nster, M\"{u}nster, Germany
\item \Idef{org55}Institut Pluridisciplinaire Hubert Curien (IPHC), Universit\'{e} de Strasbourg, CNRS-IN2P3, Strasbourg, France
\item \Idef{org56}Institute for Nuclear Research, Academy of Sciences, Moscow, Russia
\item \Idef{org57}Institute for Subatomic Physics of Utrecht University, Utrecht, Netherlands
\item \Idef{org58}Institute for Theoretical and Experimental Physics, Moscow, Russia
\item \Idef{org59}Institute of Experimental Physics, Slovak Academy of Sciences, Ko\v{s}ice, Slovakia
\item \Idef{org60}Institute of Physics, Academy of Sciences of the Czech Republic, Prague, Czech Republic
\item \Idef{org61}Institute of Physics, Bhubaneswar, India
\item \Idef{org62}Institute of Space Science (ISS), Bucharest, Romania
\item \Idef{org63}Instituto de Ciencias Nucleares, Universidad Nacional Aut\'{o}noma de M\'{e}xico, Mexico City, Mexico
\item \Idef{org64}Instituto de F\'{\i}sica, Universidad Nacional Aut\'{o}noma de M\'{e}xico, Mexico City, Mexico
\item \Idef{org65}iThemba LABS, National Research Foundation, Somerset West, South Africa
\item \Idef{org66}Joint Institute for Nuclear Research (JINR), Dubna, Russia
\item \Idef{org67}Konkuk University, Seoul, South Korea
\item \Idef{org68}Korea Institute of Science and Technology Information, Daejeon, South Korea
\item \Idef{org69}KTO Karatay University, Konya, Turkey
\item \Idef{org70}Laboratoire de Physique Corpusculaire (LPC), Clermont Universit\'{e}, Universit\'{e} Blaise Pascal, CNRS--IN2P3, Clermont-Ferrand, France
\item \Idef{org71}Laboratoire de Physique Subatomique et de Cosmologie, Universit\'{e} Grenoble-Alpes, CNRS-IN2P3, Grenoble, France
\item \Idef{org72}Laboratori Nazionali di Frascati, INFN, Frascati, Italy
\item \Idef{org73}Laboratori Nazionali di Legnaro, INFN, Legnaro, Italy
\item \Idef{org74}Lawrence Berkeley National Laboratory, Berkeley, California, United States
\item \Idef{org75}Moscow Engineering Physics Institute, Moscow, Russia
\item \Idef{org76}Nagasaki Institute of Applied Science, Nagasaki, Japan
\item \Idef{org77}National Centre for Nuclear Studies, Warsaw, Poland
\item \Idef{org78}National Institute for Physics and Nuclear Engineering, Bucharest, Romania
\item \Idef{org79}National Institute of Science Education and Research, Bhubaneswar, India
\item \Idef{org80}National Research Centre Kurchatov Institute, Moscow, Russia
\item \Idef{org81}Niels Bohr Institute, University of Copenhagen, Copenhagen, Denmark
\item \Idef{org82}Nikhef, Nationaal instituut voor subatomaire fysica, Amsterdam, Netherlands
\item \Idef{org83}Nuclear Physics Group, STFC Daresbury Laboratory, Daresbury, United Kingdom
\item \Idef{org84}Nuclear Physics Institute, Academy of Sciences of the Czech Republic, \v{R}e\v{z} u Prahy, Czech Republic
\item \Idef{org85}Oak Ridge National Laboratory, Oak Ridge, Tennessee, United States
\item \Idef{org86}Petersburg Nuclear Physics Institute, Gatchina, Russia
\item \Idef{org87}Physics Department, Creighton University, Omaha, Nebraska, United States
\item \Idef{org88}Physics Department, Panjab University, Chandigarh, India
\item \Idef{org89}Physics Department, University of Athens, Athens, Greece
\item \Idef{org90}Physics Department, University of Cape Town, Cape Town, South Africa
\item \Idef{org91}Physics Department, University of Jammu, Jammu, India
\item \Idef{org92}Physics Department, University of Rajasthan, Jaipur, India
\item \Idef{org93}Physik Department, Technische Universit\"{a}t M\"{u}nchen, Munich, Germany
\item \Idef{org94}Physikalisches Institut, Ruprecht-Karls-Universit\"{a}t Heidelberg, Heidelberg, Germany
\item \Idef{org95}Purdue University, West Lafayette, Indiana, United States
\item \Idef{org96}Pusan National University, Pusan, South Korea
\item \Idef{org97}Research Division and ExtreMe Matter Institute EMMI, GSI Helmholtzzentrum f\"ur Schwerionenforschung, Darmstadt, Germany
\item \Idef{org98}Rudjer Bo\v{s}kovi\'{c} Institute, Zagreb, Croatia
\item \Idef{org99}Russian Federal Nuclear Center (VNIIEF), Sarov, Russia
\item \Idef{org100}Saha Institute of Nuclear Physics, Kolkata, India
\item \Idef{org101}School of Physics and Astronomy, University of Birmingham, Birmingham, United Kingdom
\item \Idef{org102}Secci\'{o}n F\'{\i}sica, Departamento de Ciencias, Pontificia Universidad Cat\'{o}lica del Per\'{u}, Lima, Peru
\item \Idef{org103}Sezione INFN, Bari, Italy
\item \Idef{org104}Sezione INFN, Bologna, Italy
\item \Idef{org105}Sezione INFN, Cagliari, Italy
\item \Idef{org106}Sezione INFN, Catania, Italy
\item \Idef{org107}Sezione INFN, Padova, Italy
\item \Idef{org108}Sezione INFN, Rome, Italy
\item \Idef{org109}Sezione INFN, Trieste, Italy
\item \Idef{org110}Sezione INFN, Turin, Italy
\item \Idef{org111}SSC IHEP of NRC Kurchatov institute, Protvino, Russia
\item \Idef{org112}Stefan Meyer Institut f\"{u}r Subatomare Physik (SMI), Vienna, Austria
\item \Idef{org113}SUBATECH, Ecole des Mines de Nantes, Universit\'{e} de Nantes, CNRS-IN2P3, Nantes, France
\item \Idef{org114}Suranaree University of Technology, Nakhon Ratchasima, Thailand
\item \Idef{org115}Technical University of Ko\v{s}ice, Ko\v{s}ice, Slovakia
\item \Idef{org116}Technical University of Split FESB, Split, Croatia
\item \Idef{org117}The Henryk Niewodniczanski Institute of Nuclear Physics, Polish Academy of Sciences, Cracow, Poland
\item \Idef{org118}The University of Texas at Austin, Physics Department, Austin, Texas, USA
\item \Idef{org119}Universidad Aut\'{o}noma de Sinaloa, Culiac\'{a}n, Mexico
\item \Idef{org120}Universidade de S\~{a}o Paulo (USP), S\~{a}o Paulo, Brazil
\item \Idef{org121}Universidade Estadual de Campinas (UNICAMP), Campinas, Brazil
\item \Idef{org122}University of Houston, Houston, Texas, United States
\item \Idef{org123}University of Jyv\"{a}skyl\"{a}, Jyv\"{a}skyl\"{a}, Finland
\item \Idef{org124}University of Liverpool, Liverpool, United Kingdom
\item \Idef{org125}University of Tennessee, Knoxville, Tennessee, United States
\item \Idef{org126}University of the Witwatersrand, Johannesburg, South Africa
\item \Idef{org127}University of Tokyo, Tokyo, Japan
\item \Idef{org128}University of Tsukuba, Tsukuba, Japan
\item \Idef{org129}University of Zagreb, Zagreb, Croatia
\item \Idef{org130}Universit\'{e} de Lyon, Universit\'{e} Lyon 1, CNRS/IN2P3, IPN-Lyon, Villeurbanne, France
\item \Idef{org131}V.~Fock Institute for Physics, St. Petersburg State University, St. Petersburg, Russia
\item \Idef{org132}Variable Energy Cyclotron Centre, Kolkata, India
\item \Idef{org133}Warsaw University of Technology, Warsaw, Poland
\item \Idef{org134}Wayne State University, Detroit, Michigan, United States
\item \Idef{org135}Wigner Research Centre for Physics, Hungarian Academy of Sciences, Budapest, Hungary
\item \Idef{org136}Yale University, New Haven, Connecticut, United States
\item \Idef{org137}Yonsei University, Seoul, South Korea
\item \Idef{org138}Zentrum f\"{u}r Technologietransfer und Telekommunikation (ZTT), Fachhochschule Worms, Worms, Germany
\end{Authlist}
\endgroup

%% file: msppb-preprint.bbl
\providecommand{\href}[2]{#2}\begingroup\raggedright\begin{thebibliography}{10}

\bibitem{PhysRevLett.48.1066}
J.~Rafelski and B.~M\"uller, ``Strangeness production in the quark-gluon
  plasma,'' \href{http://dx.doi.org/10.1103/PhysRevLett.48.1066,
  10.1103/PhysRevLett.56.2334}{{\em Phys. Rev. Lett.} {\bfseries 48} (1982)
  1066--1069}. \url{http://link.aps.org/doi/10.1103/PhysRevLett.48.1066}.
  [Erratum Phys. Rev. Lett. {\bf56} (1986) 2334-2334].

\bibitem{Andersen1998209}
{\bfseries WA97} Collaboration, E.~Andersen {\em et~al.}, ``Enhancement of
  central {$\varLambda$, $\varXi$ and $\varOmega$} yields in {Pb-Pb} collisions
  at 158 {A GeV/$c$},''
  \href{http://dx.doi.org/http://dx.doi.org/10.1016/S0370-2693(98)00689-3}{{\em
  Physics Letters B} {\bfseries 433} (1998) 209 -- 216}.
  \url{http://www.sciencedirect.com/science/article/pii/S0370269398006893}.

\bibitem{Andersen1999401}
{\bfseries WA97} Collaboration, E.~Andersen {\em et~al.}, ``Strangeness
  enhancement at mid-rapidity in {Pb--Pb} collisions at 158 {$A$} {GeV}/$c$,''
  \href{http://dx.doi.org/http://dx.doi.org/10.1016/S0370-2693(99)00140-9}{{\em
  Physics Letters B} {\bfseries 449} (1999) 401 -- 406}.
  \url{http://www.sciencedirect.com/science/article/pii/S0370269399001409}.

\bibitem{Afanasiev2002275}
{\bfseries NA49} Collaboration, S.~Afanasiev {\em et~al.}, ``{$\Xi^{-}$} and
  {$\overline{\Xi}^{+}$} production in central {Pb+Pb} collisions at 158
  {GeV/$c$} per nucleon,''
  \href{http://dx.doi.org/10.1016/S0370-2693(02)01970-6}{{\em Physics Letters
  B} {\bfseries 538} (2002) 275 -- 281}.
  \url{http://www.sciencedirect.com/science/article/pii/S0370269302019706}.

\bibitem{Antinori200468}
{\bfseries NA57} Collaboration, F.~Antinori {\em et~al.}, ``Energy dependence
  of hyperon production in nucleus--nucleus collisions at {SPS},''
  \href{http://dx.doi.org/http://dx.doi.org/10.1016/j.physletb.2004.05.025}{{\em
  Physics Letters B} {\bfseries 595} (2004) 68 -- 74}.
  \url{http://www.sciencedirect.com/science/article/pii/S0370269304007725}.

\bibitem{PhysRevLett.93.022302}
{\bfseries NA49} Collaboration, T.~Anticic {\em et~al.}, ``{$\Lambda$} and
  $\overline{\Lambda}$ production in central {Pb-Pb} collisions at 40, 80, and
  {$158A\text{\,}\text{\,}\mathrm{G}\mathrm{e}\mathrm{V}$},''
  \href{http://dx.doi.org/10.1103/PhysRevLett.93.022302}{{\em Phys. Rev. Lett.}
  {\bfseries 93} (2004) 022302}.
  \url{http://link.aps.org/doi/10.1103/PhysRevLett.93.022302}.

\bibitem{PhysRevLett.92.182301}
{\bfseries STAR} Collaboration, J.~Adams {\em et~al.}, ``Multistrange baryon
  production in {Au-Au} collisions at
  $\sqrt{{s}_{NN}}=130\text{\,\hskip0.056em\hskip0.056em}\mathrm{GeV}$,''
  \href{http://dx.doi.org/10.1103/PhysRevLett.92.182301}{{\em Phys. Rev. Lett.}
  {\bfseries 92} (2004) 182301}.
  \url{http://link.aps.org/doi/10.1103/PhysRevLett.92.182301}.

\bibitem{PhysRevLett.98.062301}
{\bfseries STAR} Collaboration, J.~Adams {\em et~al.}, ``Scaling properties of
  hyperon production in $\mathrm{Au}+\mathrm{Au}$ collisions at
  $\sqrt{{s}_{NN}}=200\text{\,}\text{\,}\mathrm{GeV}$,''
  \href{http://dx.doi.org/10.1103/PhysRevLett.98.062301}{{\em Phys. Rev. Lett.}
  {\bfseries 98} (2007) 062301}.
  \url{http://link.aps.org/doi/10.1103/PhysRevLett.98.062301}.

\bibitem{PhysRevC.77.044908}
{\bfseries STAR} Collaboration, B.~I. Abelev {\em et~al.}, ``Enhanced strange
  baryon production in {Au+Au} collisions compared to $p+p$ at
  $\sqrt{{s}_{\mathit{NN}}}=200$ {GeV},''
  \href{http://dx.doi.org/10.1103/PhysRevC.77.044908}{{\em Phys. Rev. C}
  {\bfseries 77} (2008) 044908}.
  \url{http://link.aps.org/doi/10.1103/PhysRevC.77.044908}.

\bibitem{PbPbmsALICE}
{\bfseries ALICE} Collaboration, B.~Abelev {\em et~al.}, ``Multi-strange baryon
  production at mid-rapidity in {Pb--Pb} collisions at $\sqrt{s_{NN}}$ = 2.76
  {TeV},'' \href{http://dx.doi.org/10.1016/j.physletb.2013.11.048}{{\em Physics
  Letters B} {\bfseries 728} (2014) 216 -- 227}.
  \url{http://www.sciencedirect.com/science/article/pii/S0370269313009544}.

\bibitem{Redlich:2002dm}
K.~Redlich and A.~Tounsi, ``Strangeness enhancement and energy dependence in
  heavy ion collisions,''
  \href{http://dx.doi.org/10.1007/s10052-002-0983-1}{{\em The European Physical
  Journal C} {\bfseries 24} (2002) 589--594}.
  \url{http://link.springer.com/article/10.1007/s10052-002-0983-1}.

\bibitem{Kraus:2009iv}
I.~Kraus, J.~Cleymans, H.~Oeschler, and K.~Redlich, ``Particle production in
  p-p collisions and predictions for $\sqrt{s}$ = 14 {TeV} at the {CERN} {Large
  Hadron Collider (LHC)},''
  \href{http://dx.doi.org/10.1103/PhysRevC.79.014901}{{\em Phys. Rev. C}
  {\bfseries 79} (2009) 014901}.
  \url{http://link.aps.org/doi/10.1103/PhysRevC.79.014901}.

\bibitem{Becattini:2008vj}
F.~Becattini and J.~Manninen, ``Strangeness production from {SPS} to {LHC},''
  \href{http://dx.doi.org/10.1088/0954-3899/35/10/104013}{{\em Journal of
  Physics G: Nuclear and Particle Physics} {\bfseries 35} (2008) 104013}.
  \url{http://stacks.iop.org/0954-3899/35/i=10/a=104013}.

\bibitem{Aichelin:2008mi}
J.~Aichelin and K.~Werner, ``{Centrality Dependence of Strangeness Enhancement
  in Ultrarelativistic Heavy Ion Collisions: A Core-Corona Effect},''
  \href{http://dx.doi.org/10.1103/PhysRevC.79.064907,
  10.1103/PhysRevC.81.029902}{{\em Phys. Rev.} {\bfseries C79} (2009) 064907},
  \href{http://arxiv.org/abs/0810.4465}{{\ttfamily arXiv:0810.4465 [nucl-th]}}.
  \url{http://link.aps.org/doi/10.1103/PhysRevC.79.064907}.
[Erratum: Phys. Rev. C81,029902(2010)].

\bibitem{Agakishiev:2012jv}
{\bfseries STAR} Collaboration, G.~Agakishiev {\em et~al.}, ``Strangeness
  enhancement in {Cu-Cu} and {Au-Au} collisions at $\sqrt{s_{NN}}$ = 200
  {GeV},'' \href{http://dx.doi.org/10.1103/PhysRevLett.108.072301}{{\em Phys.
  Rev. Lett.} {\bfseries 108} (2012) 072301}.
  \url{http://link.aps.org/doi/10.1103/PhysRevLett.108.072301}.

\bibitem{Schnedermann:1993ws}
E.~Schnedermann, J.~Sollfrank, and U.~W. Heinz, ``Thermal phenomenology of
  hadrons from {200$A$ GeV} {S+S} collisions,''
  \href{http://dx.doi.org/10.1103/PhysRevC.48.2462}{{\em Phys. Rev. C}
  {\bfseries 48} (1993) 2462--2475},
  \href{http://arxiv.org/abs/nucl-th/9307020}{{\ttfamily arXiv:nucl-th/9307020
  [nucl-th]}}.
\url{http://link.aps.org/doi/10.1103/PhysRevC.48.2462}.

\bibitem{PhysRevC.90.054912}
V.~Begun, W.~Florkowski, and M.~Rybczynski, ``Transverse-momentum spectra of
  strange particles produced in {Pb + Pb} collisions at $\sqrt{{s}_{NN}}=2.76$
  {TeV} in the chemical nonequilibrium model,''
  \href{http://dx.doi.org/10.1103/PhysRevC.90.054912}{{\em Phys. Rev. C}
  {\bfseries 90} (2014) 054912}.
  \url{http://link.aps.org/doi/10.1103/PhysRevC.90.054912}.

\bibitem{Melo:2015iva}
I.~Melo and B.~Tomasik, ``Blast wave fits with resonances to $p_t$ spectra from
  nuclear collisions at the {LHC},'' in {\em {15th International Conference on
  Strangeness in Quark Matter (SQM 2015) Dubna, Moscow region, Russia, July
  6-11, 2015}}.
\newblock 2015.
\newblock
\href{http://arxiv.org/abs/1509.05383}{{\ttfamily arXiv:1509.05383 [nucl-th]}}.
\newblock

\bibitem{Adams2005102}
{\bfseries STAR} Collaboration, J.~Adams {\em et~al.}, ``Experimental and
  theoretical challenges in the search for the quark--gluon plasma: The {STAR}
  collaboration's critical assessment of the evidence from {RHIC} collisions,''
  \href{http://dx.doi.org/http://dx.doi.org/10.1016/j.nuclphysa.2005.03.085}{{\em
  Nuclear Physics A} {\bfseries 757} (2005) 102--183},
  \href{http://arxiv.org/abs/nucl-ex/0501009}{{\ttfamily arXiv:nucl-ex/0501009
  [nucl-ex]}}.
  \url{http://www.sciencedirect.com/science/article/pii/S0375947405005294}.

\bibitem{1748-0221-3-08-S08002}
{\bfseries ALICE} Collaboration, ``The {ALICE experiment} at the {CERN LHC},''
  {\em Journal of Instrumentation} {\bfseries 3} (2008) S08002.
  \url{http://stacks.iop.org/1748-0221/3/i=08/a=S08002}.

\bibitem{1748-0221-8-10-P10016}
{\bfseries ALICE} Collaboration, ``Performance of the {ALICE} {VZERO} system,''
  {\em Journal of Instrumentation} {\bfseries 8} (2013) P10016.
  \url{http://stacks.iop.org/1748-0221/8/i=10/a=P10016}.

\bibitem{ALICE_PseudoRapDensitypPb}
{\bfseries ALICE} Collaboration, B.~Abelev {\em et~al.}, ``Pseudorapidity
  density of charged particles in $p$ + {Pb} collisions at $\sqrt{s_{NN}}=5.02$
  {TeV},'' \href{http://dx.doi.org/10.1103/PhysRevLett.110.032301}{{\em Phys.
  Rev. Lett.} {\bfseries 110} (2013) 032301}.
\url{http://link.aps.org/doi/10.1103/PhysRevLett.110.032301}.

\bibitem{piKpLambda_pPb}
{\bfseries ALICE} Collaboration, J.~Adam {\em et~al.}, ``Multiplicity
  dependence of pion, kaon, proton and lambda production in {p-Pb} collisions
  at $\sqrt{s_{NN}}$ = 5.02 {TeV},''
  \href{http://dx.doi.org/10.1016/j.physletb.2013.11.020}{{\em Physics Letters
  B} {\bfseries 728} (2014) 25--38}.
\url{http://www.sciencedirect.com/science/article/pii/S0370269313009234}.

\bibitem{Agashe:2014kda}
Olive, K.A. $\emph{et al.}$ (Particle Data Group), Chinese Physics
  C$\textbf{38}$, 090001 (2014).
\newblock \url{http://stacks.iop.org/1674-1137/38/i=9/a=090001}.

\bibitem{ms7ppALICE}
{\bfseries ALICE} Collaboration, B.~Abelev {\em et~al.}, ``Multi-strange baryon
  production in pp collisions at $\sqrt{s}$ = 7 {TeV} with {ALICE},''
  \href{http://dx.doi.org/10.1016/j.physletb.2012.05.011}{{\em Physics Letters
  B} {\bfseries 712} (2012) 309--318}.
  \url{http://www.sciencedirect.com/science/article/pii/S037026931200528X}.

\bibitem{TheALICECollaboration:2014kr}
{\bfseries ALICE} Collaboration, ``Performance of the {ALICE} experiment at the
  {CERN LHC},'' \href{http://dx.doi.org/10.1142/S0217751X14300440}{{\em
  International Journal of Modern Physics A} {\bfseries 29} (2014) 1430044}.
  \url{http://www.worldscientific.com/doi/abs/10.1142/S0217751X14300440}.

\bibitem{DPMJet}
S.~Roesler, R.~Engel, and J.~Ranft,
  \href{http://dx.doi.org/10.1007/978-3-642-18211-2_166}{``The {Monte Carlo}
  event generator {DPMJET-III},''} in {\em Advanced Monte Carlo for Radiation
  Physics, Particle Transport Simulation and Applications}, A.~Kling,
  F.~Bar{\"a}o, M.~Nakagawa, L.~T{\'a}vora, and P.~Vaz, eds., pp.~1033--1038.
\newblock Springer Berlin Heidelberg, 2001.
\newblock \href{http://arxiv.org/abs/hep-ph/0012252}{{\ttfamily
  arXiv:hep-ph/0012252 [hep-ph]}}.
\newblock \url{http://dx.doi.org/10.1007/978-3-642-18211-2_166}.

\bibitem{Brun:1994aa}
R.~Brun, F.~Carminati, and S.~Giani,
``{GEANT Detector Description and Simulation Tool},''.

\bibitem{Tsallis:1987eu}
C.~Tsallis, ``{Possible Generalization of Boltzmann-Gibbs Statistics},''
\href{http://dx.doi.org/10.1007/BF01016429}{{\em J. Statist. Phys.} {\bfseries
  52} (1988) 479--487}.

\bibitem{Pythia6point4}
T.~Sjostrand, S.~Mrenna, and P.~Z. Skands, ``{PYTHIA 6.4 Physics and Manual},''
  \href{http://dx.doi.org/10.1088/1126-6708/2006/05/026}{{\em JHEP} {\bfseries
  05} (2006) 026},
\href{http://arxiv.org/abs/hep-ph/0603175}{{\ttfamily arXiv:hep-ph/0603175
  [hep-ph]}}.

\bibitem{PhysRevLett.111.042001}
A.~Ortiz~Velasquez, P.~Christiansen, E.~Cuautle~Flores, I.~A.
  Maldonado~Cervantes, and G.~Pai\ifmmode~\acute{c}\else \'{c}\fi{}, ``Color
  reconnection and flowlike patterns in $pp$ collisions,''
  \href{http://dx.doi.org/10.1103/PhysRevLett.111.042001}{{\em Phys. Rev.
  Lett.} {\bfseries 111} (2013) 042001}.
  \url{http://link.aps.org/doi/10.1103/PhysRevLett.111.042001}.

\bibitem{ppSpectra900ALICE}
{\bfseries ALICE} Collaboration, K.~Aamodt {\em et~al.}, ``Production of pions,
  kaons and protons in pp collisions at $\sqrt{s}$ = 900~{GeV} with {ALICE} at
  the {LHC},'' \href{http://dx.doi.org/10.1140/epjc/s10052-011-1655-9}{{\em The
  European Physical Journal C} {\bfseries 71} (2011) 1655}.
  \url{http://dx.doi.org/10.1140/epjc/s10052-011-1655-9}.

\bibitem{Strangenesspp900ALICE}
{\bfseries ALICE} Collaboration, K.~Aamodt {\em et~al.}, ``Strange particle
  production in proton-proton collisions at $\sqrt{s}$ = 0.9 {TeV} with {ALICE}
  at the {LHC},'' \href{http://dx.doi.org/10.1140/epjc/s10052-011-1594-5}{{\em
  The European Physical Journal C} {\bfseries 71} (2011) 1594}.
  \url{http://dx.doi.org/10.1140/epjc/s10052-011-1594-5}.

\bibitem{ppSpectra7TeVALICE}
{\bfseries ALICE} Collaboration, J.~Adam {\em et~al.}, ``Measurement of pion,
  kaon and proton production in proton--proton collisions at {$\sqrt{s}$} = 7
  {TeV},'' \href{http://dx.doi.org/10.1140/epjc/s10052-015-3422-9}{{\em The
  European Physical Journal C} {\bfseries 75} (2015) 226}.
  \url{http://dx.doi.org/10.1140/epjc/s10052-015-3422-9}.

\bibitem{Andronic:2009ht}
A.~Andronic, P.~Braun-Munzinger, and J.~Stachel, ``Thermal hadron production in
  relativistic nuclear collisions: The hadron mass spectrum, the horn, and the
  {QCD} phase transition,''
  \href{http://dx.doi.org/10.1016/j.physletb.2009.02.014}{{\em Physics Letters
  B} {\bfseries 673} (2009) 142--145}.
  \url{http://www.sciencedirect.com/science/article/pii/S0370269309001609}.

\bibitem{Wheaton200984}
S.~Wheaton, J.~Cleymans, and M.~Hauer, ``{THERMUS} --- a thermal model package
  for {ROOT},''
  \href{http://dx.doi.org/http://dx.doi.org/10.1016/j.cpc.2008.08.001}{{\em
  Computer Physics Communications} {\bfseries 180} (2009) 84 -- 106}.
  \url{http://www.sciencedirect.com/science/article/pii/S0010465508002750}.

\bibitem{KosLambdaPbPbALICe}
{\bfseries ALICE} Collaboration, B.~B. Abelev {\em et~al.}, ``{$K^0_S$ and
  $\Lambda$ production in Pb-Pb collisions at $\sqrt{s_{NN}}$ = 2.76 TeV},''
  \href{http://dx.doi.org/10.1103/PhysRevLett.111.222301}{{\em Phys. Rev.
  Lett.} {\bfseries 111} (2013) 222301},
\href{http://arxiv.org/abs/1307.5530}{{\ttfamily arXiv:1307.5530 [nucl-ex]}}.

\end{thebibliography}\endgroup
